\documentclass[conference]{IEEEtran}
\IEEEoverridecommandlockouts

\usepackage{cite}
\usepackage{tikz}
\usepackage{xspace}
\usepackage{amsmath,amsfonts}

\usepackage{amssymb}
\usepackage{algorithmic}
\usepackage{graphicx}
\usepackage{textcomp}
\usepackage{xcolor}
\usepackage{tabularx}
\usepackage{makecell}
\usepackage{listings}

\def\BibTeX{{\rm B\kern-.05em{\sc i\kern-.025em b}\kern-.08em
    T\kern-.1667em\lower.7ex\hbox{E}\kern-.125emX}}

\newcommand*\myorangecirc[1]{%
    \begin{tikzpicture}[baseline=(C.base)]
        \node[fill=orange,text=white,draw=orange,circle,inner sep=0.5pt](C) {#1};
    \end{tikzpicture}}

\newcommand{\code}[1]{\texttt{#1}\xspace}

\newcommand{\blue}[1]{{#1}}

\newcommand{\roboRT}{RoboCore\xspace}
\newcommand{\tensorcore}{tensor accelerator\xspace}
\newcommand{\furthestSuccessRate}{94.8\%\xspace}
\newcommand{\randomSuccessRate}{88.7\%\xspace}
\newcommand{\rtnnConfigA}{P-Ray\xspace}
\newcommand{\rtnnConfigB}{P-Sphere\xspace}
\newcommand{\speedupCollision}{2.8$\times$\xspace}
\newcommand{\speedupCollisionCuda}{13.3$\times$\xspace}
\newcommand{\speedupMpiNet}{3.4$\times$\xspace}

\newcommand{\energysavings}{48\%\xspace}
\newcommand{\areaoverhead}{2\%\xspace}

\definecolor{codegreen}{rgb}{0,0.6,0}
\definecolor{codegray}{rgb}{0.5,0.5,0.5}
\definecolor{codepurple}{rgb}{0.58,0,0.82}
\definecolor{backcolour}{rgb}{0.95,0.95,0.92}

\begin{document}

\title{RoboGPU: Accelerating GPU Collision Detection for Robotics}
\author{
  \IEEEauthorblockN{
    Lufei Liu, 
    Liwei Xue, 
    Yuan Hsi Chou, 
    Jocelyn Zhao,
    Lara Kawasme,
    Youssef Mohammed,
    Tor M. Aamodt
  }
  \IEEEauthorblockA{
    \textit{University of British Columbia} \\
    Vancouver, Canada \\
    liulufei@student.ubc.ca, aamodt@ece.ubc.ca \\ 
  }
}














\maketitle
\begin{abstract}
Autonomous robots are anticipated to be deployed soon in domains ranging from transportation to healthcare and home assistance.
Enabling autonomous robotics requires a computation platform flexible enough to execute a diverse and evolving collection of workloads while meeting real-time requirements.
We believe a GPU-like architecture will be a key component of such platforms.
Recent GPUs combine a flexible parallel processing fabric augmented with efficient support for important application domains via embedded accelerators (e.g., Tensor Cores), and a GPU-like architecture has reportedly been adopted for the Tesla AI5 accelerator.
While current GPUs are effective at supporting emerging neural motion planners, we find that collision detection is crucial for evaluating their proposed trajectories and that this step appears to require dedicated acceleration to operate in real-time. 
In this work, we propose \roboRT, an accelerator block embedded within a robotics-focused GPU (RoboGPU) architecture.
We explore and compare architectural modifications to address the gaps of existing ray tracing accelerators (RTAs) for robotics and find that \roboRT computes collision queries \speedupCollision faster than RTA implementations using \energysavings less energy with \areaoverhead more area than RTAs.
\roboRT is \speedupCollisionCuda faster than a CUDA baseline, and achieves \speedupMpiNet end-to-end speedup on a neural motion planner and 1.1$\times$ speedup on Monte Carlo Localization compared to a baseline GPU.
This demonstrates that a hybrid approach of embedded specialization within a flexible general-purpose GPU architecture is suitable for supporting advancements in robotics.
Code available at: https://ubc-aamodt-group.github.io/robogpu/
\end{abstract}


\section{Introduction}
\label{sec:introduction}

Autonomous robots have significant potential across industries, including self-driving vehicles, healthcare, and home assistance.
Recent advances in neural motion planning, both in path planning with MPNet~\cite{mpnet} and M$\pi$Net~\cite{mpinet} and vision-language-action (VLA) models such as NVIDIA's Alpamayo-R1~\cite{alpamayo} have brought the development of truly autonomous robots closer to reality. 
These models help robots interpret environments, identify goals, and compute feasible trajectories to follow.
Companies such as Tesla, Waymo, and Mercedes are already shifting from classical algorithms to neural models in self-driving vehicles and humanoid robots~\cite{ashok2025,miskovetz2024actuator, ogale2021neural,soares2026mercedes_sclass_drive_av}, and we speculate that this trend will continue as more robots make the transition.
However, despite a growing number of startups, some roboticists believe we may still be a decade from widespread robotics deployment in everyday life~\cite{brooks2024three_laws_robotics}. 
Thus, robotics algorithms will likely continue to evolve and consequently need a flexible yet efficient compute platforms to handle the real-time compute-intensive tasks robotics requires.

\begin{figure}[t]
    \centering
    \includegraphics[width=0.5\textwidth]{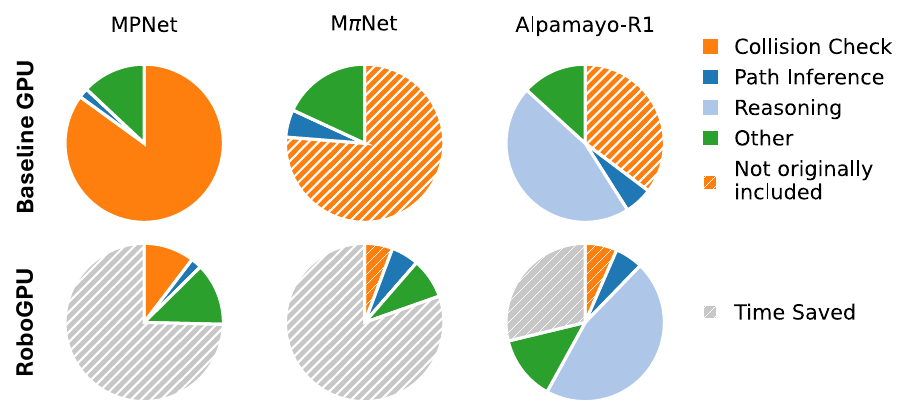}
    \caption{Latency distribution of neural planning on MPNet, M$\pi$Net, and Alpamayo-R1 with explicit collision checking (appended for M$\pi$Net and Alpamayo-R1).}
    \label{fig:neural-planner-collision}
\end{figure}

Moreover, despite recent success rate improvements of published neural motion planners, empirical results demonstrate these models remain imperfect. 
Thus, similar to Qureshi et al.~\cite{mpnet}, we assume that future systems deploying neural planners autonomously in safety critical environments where actual collisions are unacceptable will couple neural path planning with explicit collision detection~\cite{mpinet,fishman2024avoid}.  
This is akin to the philosophy underlying DIVA~\cite{austin1999diva}, in which a validated in-order checker-processor can quickly verify the correctness of already  executed instructions that are about to be committed by a fast, possibly buggy, out-of-order core since dependencies are decoupled post-execution.  
Since verifying the behavior of neural networks is challenging~\cite{dutta2018output}, we study a system in which a path is found using an approximate neural planner (not proven to guarantee safety) and the proposed path is checked using a more conservative classical collision detection algorithm before control outputs would be sent to physical actuators.

While recently proposed neural planners are trained with collision-avoidance objectives, they are known to still produce paths that collide with obstacles in the environment. 
For example, M$\pi$Net reports 99\% success in finding a path reaching the target goal but the proposed paths result in a 14\% collision rate, bringing the actual collision-free success rate to just 80.80\%~\cite{fishman2024avoid}.
AvoidEverything~\cite{fishman2024avoid} results in a 3.10\% collision rate despite optimizing specifically for collision avoidance leading to a 92.34\% success rate. 
More recently, Alpamayo-R1~\cite{alpamayo} reports a 9\% close encounter rate for their 0.5B real-time model, which is non-negligible and likely beyond reasonable incidents per million miles (IPMM) metrics for self-driving vehicles~\cite{kusano2024comparison}. 
Avoiding collisions using neural models becomes more difficult for humanoid robots with higher degrees of freedom~\cite{guo2025deepcollide} that cannot easily recover from instability or falling.
Dedicated collision avoidance units exist in many robots already~\cite{ogale2021neural, nister2021safety, floyd2025collision}. 
Therefore, consistent with limitations identified in the neural path planning literature~\cite{mpinet,fishman2024avoid}, we believe explicit collision detection will be crucial in the robotics pipeline when adopting neural motion planners for safety-critical tasks. 

However, collision detection is known to be costly both in time and energy~\cite{murray2016microarchitecture,shah2024collision,zhao2025neural}.
Figure~\ref{fig:neural-planner-collision} shows the latency distribution of neural planning with MPNet and M$\pi$Net measured on a NVIDIA RTX 5070 GPU, and Alpamayo-R1 using reported results \blue{(top)}.
MPNet already includes collision checking and we add collision checking to M$\pi$Net and Alpamayo-R1 to estimate the latency impact.
Although collision detection is a smaller component in Alpamayo-R1 where reasoning latency dominates, reasoning only produces the target goal, while path planning and collision checking are repeatedly executed for replanning.
Collision queries add an additional 76\% latency to low level planners like M$\pi$Net and remain a bottleneck at 65\% of replanning latency for a VLA model.
\blue{Figure~\ref{fig:neural-planner-collision} (bottom) shows how a dedicated robotics architecture, RoboGPU, can reduce the overall latency while ensuring safe collision checking, ultimately resulting in improved robotic responsiveness~\cite{leven2002framework}.}

Many specialized accelerators have been proposed to meet the latency requirements of robotics tasks~\cite{sorin,daducd,racod,wan2025reca,huang2024moped}, but they are limited to specific algorithms and cannot adapt to new robotics advancements.
The number of papers accepted at ICRA, the leading robotics conference, has roughly doubled in the past ten years~\cite{icra_papers}, showing robotics algorithms are evolving rapidly.  
To leverage the benefits, there is a need to combine programmability with efficiency in robotics computation platforms. 
We speculate that system-on-chips integrating general-purpose Graphics Processing Units (GPUs) are strong candidates for robotics acceleration due to their existing combination of programmability and high computational performance.
For example, NVIDIA's Jetson GPUs are designed with the flexibility to target robotics applications across a range of domains and Tesla  recently indicated AI5, their latest  chip, is a GPU-like architecture optimized for  autonomous vehicles\footnote{Elon Musk, describing deletion of the ``legacy GPU'' in the upcoming AI5 self-driving chip, suggested this was motivated ``because [AI5] basically {\em is} a GPU'' for which ``excess production can always be put in our data centers [alongside NVIDIA GPUs].''~\cite{Tesla2025Q3EarningsQA}}.

\begin{figure}[t]
    \centering
    \includegraphics[width=0.5\textwidth]{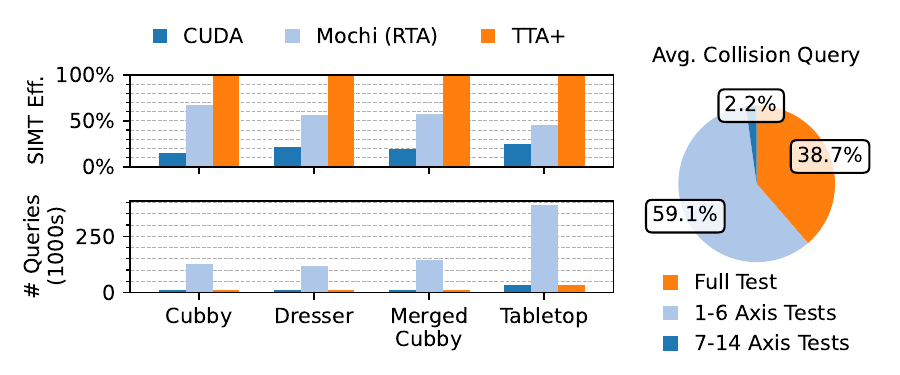}
    \caption{SIMT efficiency and collision queries required by each device on environments from M$\pi$Net~\cite{mpinet} (left). Required computation for the average collision query (right).}
    \label{fig:profile}
\end{figure}

Today's GPUs increasingly incorporate hardened accelerator blocks such as Tensor Cores, Tensor Memory Accelerators, and Ray Tracing Cores to improve performance and efficiency without sacrificing flexibility, similar to how embedding ``hard'' multiplier and memory blocks were adopted within FPGA fabrics two decades ago to help close area, performance, and power gaps between completely flexible lookup tables and fixed function ASICs~\cite{kuon2006measuring}. 
Despite strong competition~\cite{GoogleTPU}, NVIDIA has repeatedly shown that the tension between programmability and efficiency can be balanced in part by integrating such hardware units into the general-purpose GPU architecture united by the CUDA programming model. 
This approach to balancing efficiency and flexibility is also consistent with earlier academic studies~\cite{hameed2010understanding}.
We speculate that with adoption of neural techniques, adding such a dedicated robotics accelerator block for collision detection will be increasingly helpful for ensuring safety, especially as software collision detection struggles to meet real-time latency constraints.
Figure~\ref{fig:profile} (top left) highlights this issue appearing as low SIMT efficiency (threads active per warp due to control flow) for CUDA collision queries on a NVIDIA RTX 2080 Ti GPU, causing underutilization.

We explore collision detection using a dedicated hardware unit \textit{within} the GPU to achieve real-time performance, as offloading queries to the CPU or other accelerators would incur severe communication overheads and latency, especially considering recent trends of CPU orchestration bottlenecks~\cite{raj2025cpu}.
A close match to the hardware requirements is the Ray Tracing Accelerator (RTA) in current GPUs~\cite{rtcore} since the ray tracing operation is also divergent and generates irregular memory accesses in 3D space, with several works already attempting to leverage RTAs for robotics workloads~\cite{mandarapu2024mochi,min2023octomap,sui2025hardware}.
Figure~\ref{fig:profile} shows that collision queries on the RTA using Mochi~\cite{mandarapu2024mochi} hides the divergence better than CUDA but requires 12$\times$ more queries in its software-only implementation for the same task due to the RTA interface.
Other generalized versions of RTAs, such as TTA$+$~\cite{ha2024tta}, can achieve higher SIMT efficiency without the overheads of regular RTAs but suffer different inefficiencies for collision queries.
Although RTAs are not ideal for robotics, using RTAs as the foundation for \roboRT allows for minimal design costs and leverage of existing hardware and software infrastructure.

In this paper, we propose \roboRT, a hardware unit that integrates into existing GPU architectures \blue{to create RoboGPU} and provide an RTA-like platform optimized for robotics workloads.
We focus our design on accelerating explicit collision detection \blue{as a step towards safe deployment of neural motion planners in real-world robotics}.
By placing \roboRT within the GPU, we can eliminate data transfers from the neural planner to \roboRT for collision queries, tightly coupling the two stages for efficient planning iterations.
We ensure \roboRT also adapts well to other important kernels in the robotics pipeline such as point cloud processing and classical robotics algorithms such as Monte Carlo Localization, highlighting its flexibility to support advancements in robotics.
Our contributions are as follows:

\begin{itemize}
    \item We propose \blue{RoboGPU, a robotics-focused GPU that integrates our \roboRT design} to enable fast collision detection queries, achieving \speedupCollision speedup over existing RTA implementations while maintaining programmability.
    \item We compare design choices for \roboRT and analyze their impact on performance and efficiency for robotics collision detection, resulting in \energysavings energy savings with \areaoverhead area overhead over RTAs.
    \item We demonstrate that \roboRT extends beyond collision detection \blue{to a subset of robotics kernels}, accelerating point cloud processing in M$\pi$Net~\cite{mpinet} by \speedupMpiNet and Monte Carlo Localization from the RoWild~\cite{rowild} benchmark by 1.1$\times$ over a baseline GPU.
\end{itemize}
\section{Background}
\label{sec:background}

This section provides an overview of motion planning for robotics, including neural motion planners and collision detection.
We also describe details of RTAs in modern GPUs and prior works that leverage RTAs for alternative workloads.

\subsection{Motion Planning for Robotics}
\begin{figure}
    \centering
    \includegraphics[width=0.5\textwidth]{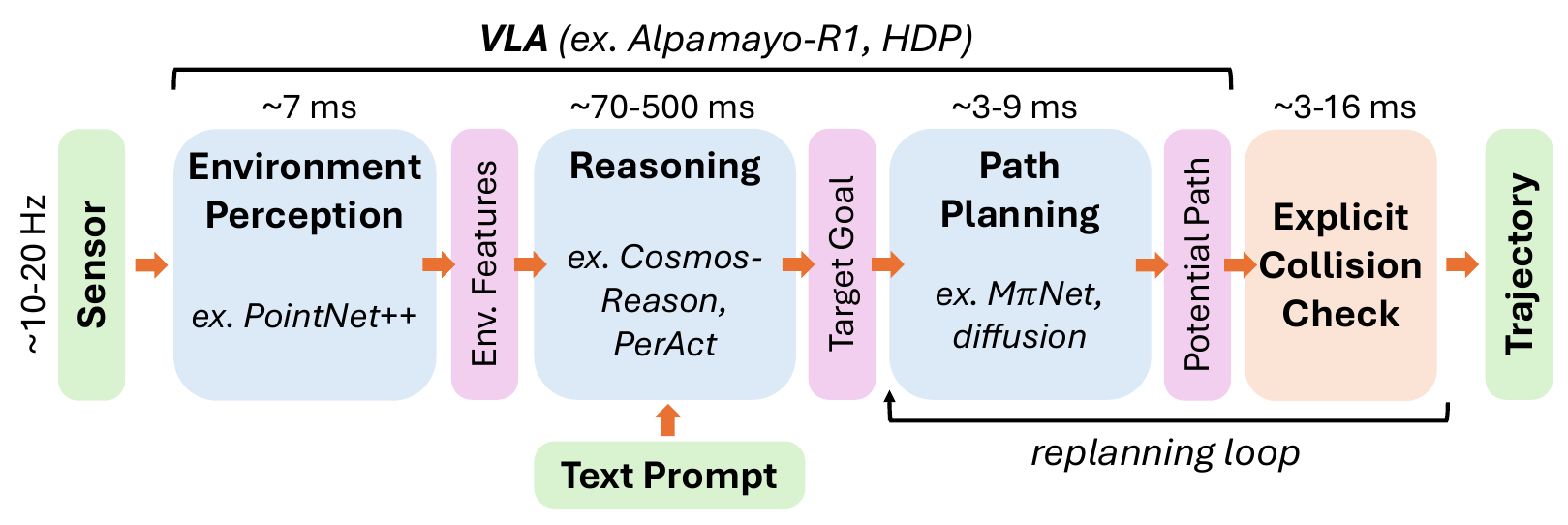}
    \caption{Robotics pipeline with VLA model and collision checking. The VLA model consists of individual stages with sensors and text as inputs and a potential path as the output.}
    \label{fig:vla}
\end{figure}

Motion planning in robotics aims to identify a feasible trajectory for a robot to move from its current position to reach a target goal, broken into a sequence of discretized milestones.
Traditionally, these milestones are generated using algorithmic methods, including search-based planning algorithms~\cite{astar1, astar2} and sampling-based algorithms~\cite{rrtstar, samplingalgo, bitstar}.
With the rise of deep learning, neural path planners such as M$\pi$Net~\cite{mpinet} and others~\cite{mpnet,dalal2024neuralmp,huang2024neural,yu2024efficient} have been introduced, seeking to imitate efficient samplers to speed up traditional planners.
These planners all require a meaningful representation of the environment, typically derived from sensor inputs such as LiDAR or RGB-D cameras that are preprocessed by other neural networks like PointNet++~\cite{pointnetplus} to extract relevant features.
More recently, vision-language-action (VLA) models such as HDP~\cite{ma2024hierarchical} and Alpamayo-R1~\cite{alpamayo} have combined these tasks into a single end-to-end model that takes sensor inputs and natural language instructions to directly output a robot path.
However, the underlying architecture of these VLA models still generally consists of an encoder to understand the environment, a reasoning module to identify the target goal, and a low-level path planner that generates trajectory milestones, summarized \blue{with latencies} in Figure~\ref{fig:vla}.
Low-level path planners are often implemented with diffusion models, inspired by Diffusion Policy~\cite{chi2025diffusion}, which are used for both Alpamayo-R1 and HDP.
Although neural motion planning networks are becoming increasingly powerful and attaining higher success rates, the failure rate increases significantly with the complexity of the environment and the difficulty of the end goal. 

\subsection{Collision Detection}
Collision detection identifies whether a robot will collide with obstacles in its environment along a path, and is crucial for safety when using neural planners~\cite{mpinet, fishman2024avoid}.
Although there are neural models for collision detection like DeepCollide~\cite{guo2025deepcollide}, we focus on classical algorithms to avoid introducing additional errors. 
Collision checks can be tightly coupled with the path planning stage in Figure~\ref{fig:vla} to select between potential paths and replan when necessary.
The replanning loop is especially time-sensitive as the robot must react to environmental changes and update its path in real-time, whereas the initial goal reasoning stage can be more flexible in latency.

To perform collision checks, the potential robot path is discretized and the robot geometry at each step can be represented by oriented bounding boxes (OBBs) for a conservative estimate. 
The environment can be represented with axis-aligned bounding boxes (AABBs).
AABBs, unlike OBBs, are aligned with the coordinate axes, making them simpler to compute and store.
These AABBs are then stored in spatial data structures, such as the octree that recursively subdivides the 3D space into octants hierarchically, or the bounding volume hierarchies (BVH) that tightly bounds groups of nearby AABBs (possibly overlapping) into larger AABBs to form a tree structure.
We evaluate octrees over BVHs since they are easier to store and update, which is more suitable for robotics, although the AABBs used for the collision algorithm are the same for both.
\blue{Octree construction and updates are a time consuming process and can potentially impact the overall performance of collision detection.}
\blue{However, we focus on accelerating the collision query stage in this work and rely on existing octree optimization strategies~\cite{deng2017g, chen2025octocache, min2023octomap,durvasula2022voxelcache} to mitigate this issue.}

\begin{figure}
    \centering
    \includegraphics[width=0.5\textwidth]{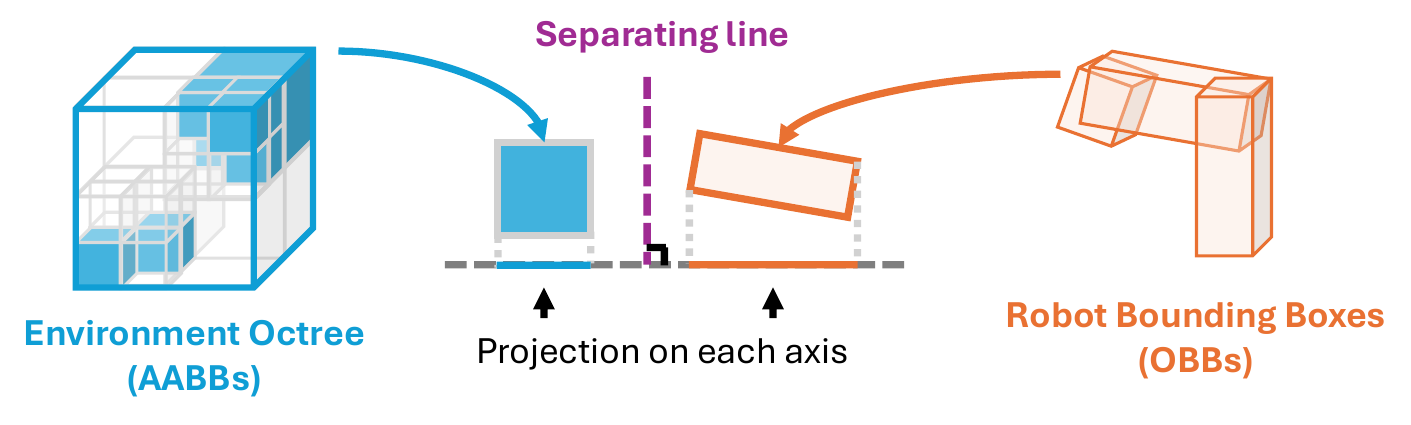}
    \caption{Separating axis collision test (SACT) between AABB from environment octree and OBB from robot. If a separating line divides the projections of the two boxes, then the boxes are not colliding. Reproduced from~\cite{energyefficientmotionplanning2024shah}.}
    \label{fig:collision_algorithm}
\end{figure}

Collision detection between the robot and environment is performed by traversing the octree and checking for intersections between the robot OBBs and the occupied environment AABBs at each step.
Figure~\ref{fig:collision_algorithm} illustrates the separating axis collision test (SACT) used for OBB-AABB intersection testing~\cite{obbtree1996Gottschalk}.
Both the AABB and OBB are projected onto potential separating axes, and if there exists a separating line on any axis that divides the projections of the two boxes, then the boxes are not colliding.
The SACT creates many early exit opportunities since finding a separating line on any axis guarantees no collision, eliminating the need to check remaining axes.
Without early exits, each test requires 30 projections and 15 overlap checks, making collision queries of many robot poses against complex environments in real-time quite challenging.

\subsection{GPU Hardware}

\begin{figure}
    \centering
    \includegraphics[width=0.49\textwidth]{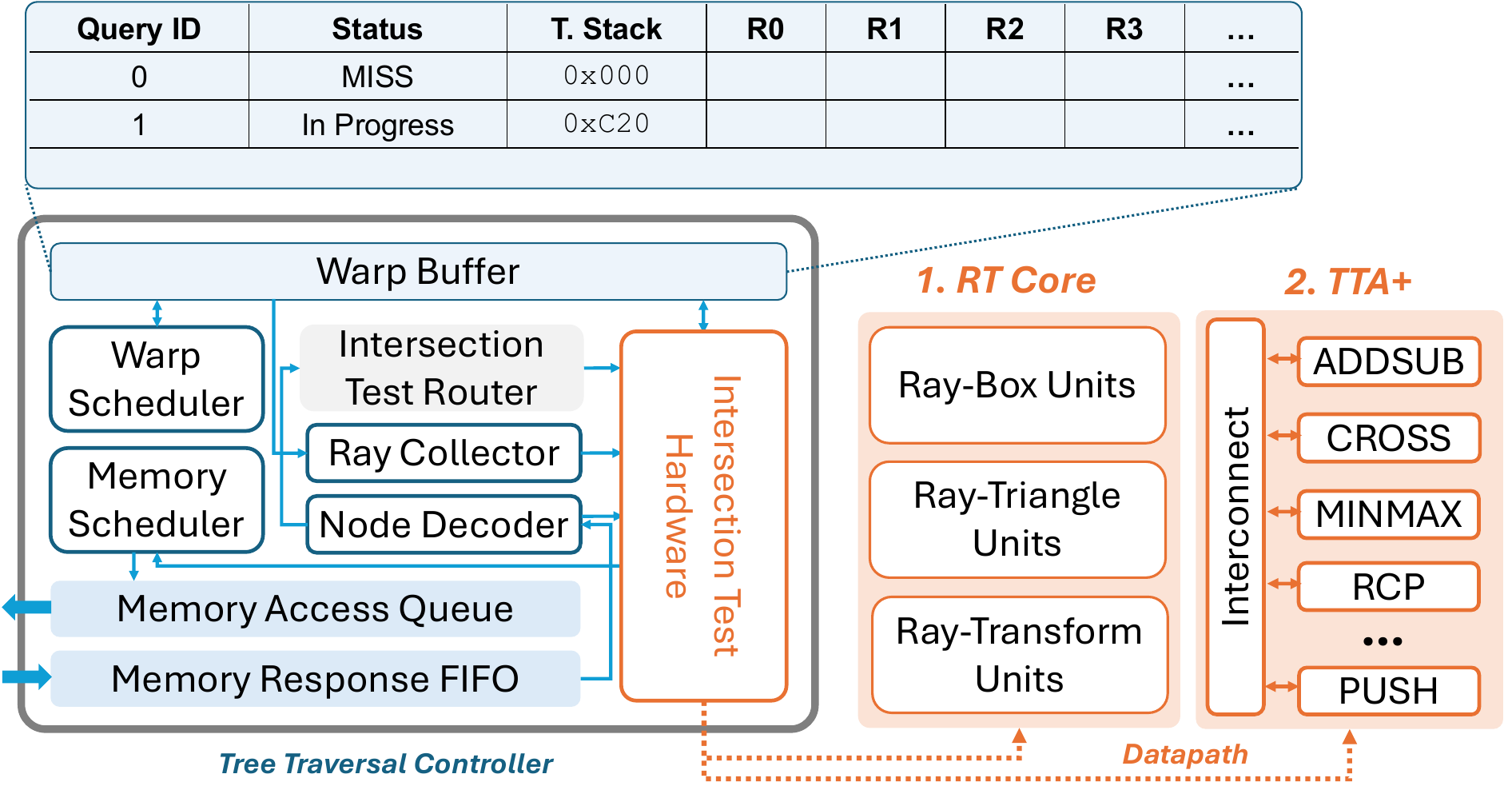}
    \caption{Ray tracing accelerator architecture on a GPU, which may include fix-function intersection hardware (RT Core design~\cite{rtcore}) or programmable operation units (TTA$+$ design~\cite{ha2024tta}) to execute intersection tests.}
    \label{fig:baselinertunit}
\end{figure}

GPUs consist of thousands of threads and are designed to accelerate massively parallel applications.
Threads in a warp are executed in Single Instruction Multiple Thread (SIMT) fashion, where the same instruction is executed on different data.
However, the SIMT architecture does not handle irregular applications well, such as ray tracing, which traverses a BVH tree and performs intersection tests between rays and geometry.
Therefore, modern GPUs include a specialized Ray Tracing Accelerator (RTA) to address these performance bottlenecks~\cite{rtcore,intel_arc,amd_rdna3,imaginationtech,apple_m3}.
GPUs also include a neural accelerator, like the Tensor Core, that efficiently perform matrix multiplications in neural network inference. 

Figure~\ref{fig:baselinertunit} shows the RTA architecture on the GPU.
RTAs can be broken down to two main components, the \textit{tree traversal controller} that coordinates with the main GPU cores and fetches required data from memory, and the intersection units forming a \textit{datapath} for computation.
When threads in a warp issue a \code{traceRay} instruction to the RTA, the per-ray information such as ray origin, direction, and a traversal stack are stored in the warp buffer.
Every cycle, the warp scheduler selects a warp to execute and issues memory requests to fetch the required tree nodes from memory.
Once the data returns, the rays and node data are routed based on the node type to the corresponding intersection units.
The intersection units can be implemented as fixed-function hardware, the typical design for ray tracing, or as programmable operation units (OP units) as proposed in the tree traversal accelerator (TTA$+$) design~\cite{ha2024tta}.
TTA$+$ targets flexibility, with a modular system that decomposes intersection units into individual OP units and connecting them through an interconnect similar to a dataflow architecture~\cite{dennis1974preliminary}.
TTA$+$ executes a custom instruction set called $\mu$ops, where each $\mu$op corresponds to a specific OP unit and an OP destination table specifies which OP unit the output data should route to next.
However, TTA$+$ was not designed for instruction sequences as complex as the SACT and has no ability to support early exits.

Although RTAs are designed for computer graphics, they are often repurposed for other applications such as FastRNN~\cite{RT_RadiusSearch} and RTNN~\cite{zhu2022rtnn} for nearest neighbor search, Mochi~\cite{mandarapu2024mochi} and RT-DCD~\cite{sui2025hardware} for collision detection, and many others~\cite{zhang2025rtspmspm,henneberg2023rtindex,nagarajan2023rt}.
These works reformulate key algorithms into the ray tracing paradigm to run on RTAs, representing data as 3D points or volumes stored into a BVH structure and issuing ``rays" to query the data.
For example, RTNN represents data points as spheres, then issues rays to find the nearest neighbors by performing ray-sphere intersection tests.
However, these reformulations cause redundant handoffs to non-accelerated intersection shaders that hinder RTA benefits.
\section{\roboRT Architecture}
\label{sec:architecture}
Our goal is to design a hardware unit for a RoboGPU that accelerates collision detection with the ability to adapt as robotics workloads inevitably evolve.
Ideally, this unit requires minimal hardware modifications to reduce design and validation costs, making it a practical implementation for GPU vendors.
Existing RTAs provide an excellent platform for tree traversals, but they are inefficient for collision detection.
This section addresses the limitations of RTA designs, including the TTA$+$, and introduces our \roboRT architecture.

\subsection{High Level Design}
\label{ssec:intersectionprogram}

\begin{figure}[t]
    \centering
    \includegraphics[width=0.4\textwidth]{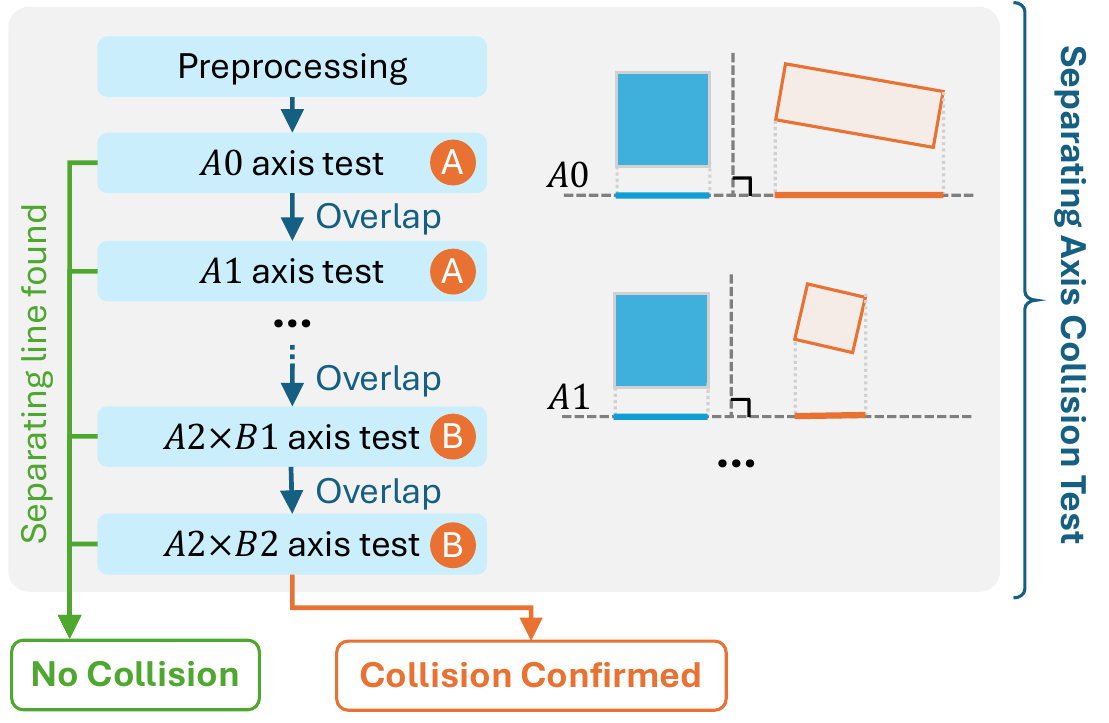}
    \caption{Staged collision detection with early termination. Box-Normal axis tests executes first, marked with (A), then Edge$\times$Edge axis tests marked with (B). Reproduced from~\cite{energyefficientmotionplanning2024shah}.}
    \label{fig:earlyexit}
\end{figure}

Figure~\ref{fig:earlyexit} illustrates the staged collision detection process that executes the 15-axis SACT, which is the main algorithm that \roboRT targets.
The SACT workload exhibits short but highly repetitive instruction sequences that occur in bursts across a large number of independent queries, which is ideal for a fixed-function accelerator.
However, robotics pipelines evolve quickly, so a purely fixed-function collision unit would be too rigid to support new emerging kernels and becomes obsolete over time.
Meanwhile, a fully programmable CPU-like core is flexible but incurs unnecessary overheads with instruction fetch and decode since SACT repeats the same operation sequence for every query.
Dataflow and CGRA-style architectures face other inefficiencies with token tracking and coordinating early exits between independent queries.
The closest existing solution that balances programmability and efficiency are RTAs, which take advantage of fixed-function datapaths but are general enough to support computer graphics as the field grows.

Programmable RTAs such as the TTA$+$ can execute SACT using the list of $\mu$ops in Table~\ref{table:sactops}.
However, the complete SACT requires a total of 47 $\mu$ops to test all 15 axes, which is significantly more than the original TTA$+$ workloads that required 12 $\mu$ops on average to 19 $\mu$ops at maximum per intersection program~\cite{ha2024tta}.
Unlike dedicated collision detection accelerators such as MPAccel~\cite{energyefficientmotionplanning2024shah}, TTA$+$ has no mechanism to skip unnecessary axes checks using early exits, which is crucial for performance and would reduce computation latency by 65\% on average per query.
\roboRT addresses this gap by specializing the intersection datapath for SACT and adding early-exit control while preserving programmable execution for non-SACT kernels.
This includes ray tracing kernels if needed since ray tracing does occasionally appear in robotics~\cite{min2023octomap, min2021accelerating}.
Outside the intersection path, \roboRT reuses the same tree traversal controller as RTAs, which integrates well into existing GPU architectures.
Rather than inserting an entirely new accelerator, reusing the proven RTA infrastructure where possible lowers the barrier of extending GPU capabilities and amortizes hardware investments through flexibility across applications as they evolve.

\begin{table}[t]
    \caption{$\mu$ops used for SACT intersection program}
    \label{table:sactops}
    \centering
    \begin{tabularx}{0.48\textwidth}{|l|l|X|}
    \hline
    \textbf{Test type} & \textbf{\# of tests} & \textbf{TTA$+$ $\mu$ops used per test} \\ \hline
    Pre-Process & 1 & 4 VEC3-SUB, 1 R-XFORM \\ \hline
    \myorangecirc{A} Box-Normal Axis & 6 & 2 DOT, 1 VEC3-SUB, 1 VEC3-CMP\\ \hline
    \myorangecirc{B} Edge$\times$Edge Axis & 9 & 1 CROSS, 1 VEC3-CMP \\ \hline
    \end{tabularx}%
\end{table}

\subsection{\roboRT Datapath}

\begin{figure*}[t]
    \centering
    \includegraphics[width=\textwidth]{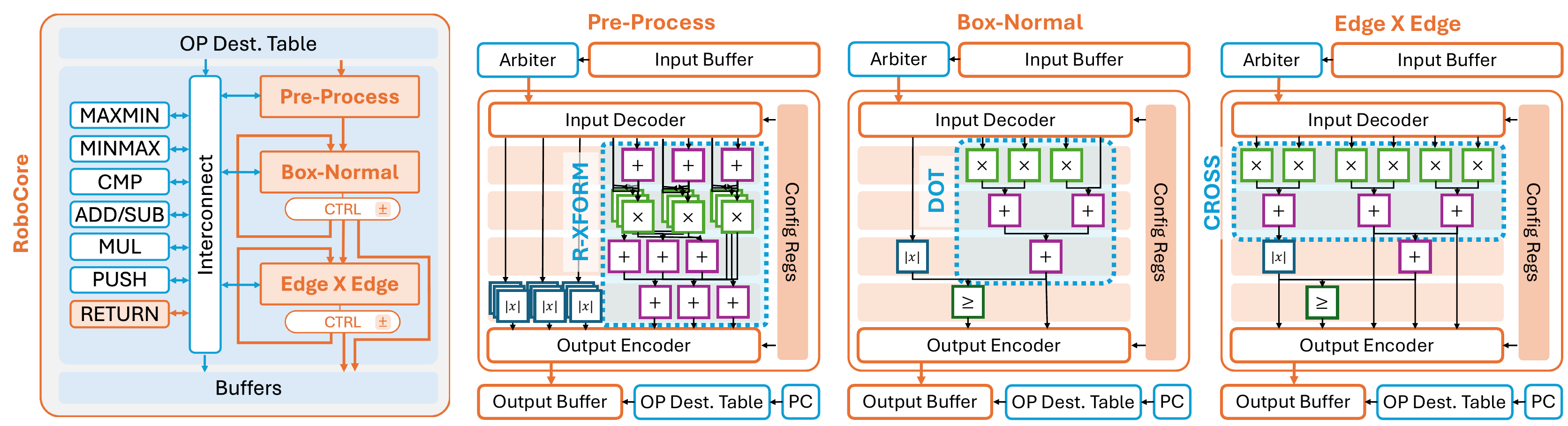}
    \caption{\roboRT datapath using three specialized SACT OP units with embedded general $\mu$ops.}
    \label{fig:robocore_units}
\end{figure*}

\roboRT adds three new four-stage pipelined SACT units (Pre-Process, Box-Normal, and Edge$\times$Edge) that merge TTA$+$ $\mu$op sequences into direct computation per test type, as listed by the SACT decomposition in Table~\ref{table:sactops}.
Figure~\ref{fig:robocore_units} shows the \roboRT datapath, which combines the new SACT units with the existing programmable TTA$+$ model.
Beyond the redesigned datapath, we reuse the baseline RTA traversal hardware for \roboRT, including node fetching, traversal-state tracking, and traversal scheduling, which are already heavily optimized.
This choice reduces integration overheads and allows \roboRT to benefit from orthogonal traversal optimizations such as treelet prefetching and more~\cite{treeletprefetch,tozlu2025cooprt,liu2021intersection}.

Each SACT unit features a compute pipeline composed of floating point arithmetic elements such as multipliers, adders, and comparators to execute the required operations for the tests in Table~\ref{table:sactops}.
Compared to the TTA$+$ implementation, these SACT units are able to condense and parallelize the operations for each test, which reduces the area overhead of the specialized hardware while also reducing the total latency per test.
Each unit still exposes an embedded general $\mu$op (R-XFORM, DOT, and CROSS), which accounts for the largest portion of the SACT units and can be reused in other robotics kernels despite the specialized design.
The SACT unit also includes the same input buffer, input decoder, output encoder, and output buffer as a standard TTA$+$ OP unit for compatibility. 
These components allow the SACT units to share the same interconnect packet format used in TTA$+$ that can be passed between units and communicate with the main traversal controller.

The SACT units are connected in series to mimic the 15-axis test sequence in Figure~\ref{fig:earlyexit}, with the Pre-Process unit feeding into the Box-Normal unit, which feeds into the Edge$\times$Edge unit.
To execute the SACT program, the traversal controller issues a packet to the Pre-Process unit, which executes the first stage of the SACT then autonomously forwards the results to the next SACT unit without going back through the interconnect, which is a source of overhead on TTA$+$.
Alternatively, to use the SACT unit as a general $\mu$op, any OP unit can issue a packet directly through the interconnect to any of the three SACT units, which can be configured to only execute the embedded $\mu$op bypassing any SACT-specific logic in the datapath.
The OP unit then sends the results back through the interconnect to the next unit based on the configuration registers.
At the datapath boundary, each unit accepts packets either from the interconnect (programmable mode) or from the previous SACT stage (specialized mode).
An input arbiter prioritizes in-flight SACT packets to preserve collision detection latency, while programmable packets may experience some delay since they can tolerate higher latency.
This dual-mode behavior lets \roboRT amortize the area of specialized hardware while keeping the same programming model and interconnect protocols for other robotics kernels.

\subsubsection{Early-Exit Control}

\begin{figure}[t]
    \centering
    \includegraphics[width=0.4\textwidth]{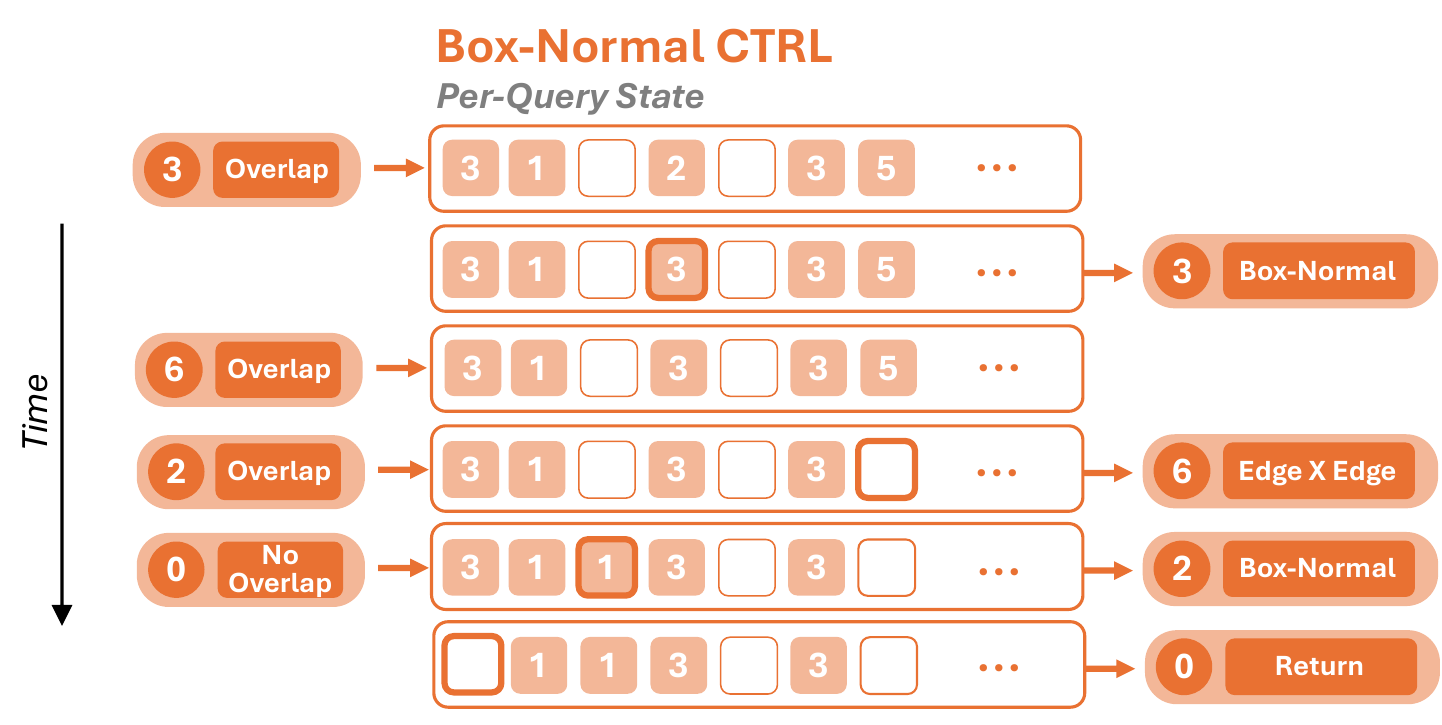}
    \caption{\blue{Example of CTRL logic for Box-Normal unit.}}
    \label{fig:robocore_ctrl}
\end{figure}

To support early exits from the SACT program, we insert control logic (CTRL) in \roboRT between the three SACT units.
The controller tracks per-query progress using \blue{query IDs} and counters for the 15 axes, which are incremented when an axis test passes and reset when a new query is issued.
The Box-Normal counters tracks the first six axes tests, while the Edge$\times$Edge counters tracks the remaining nine axes.
The controller monitors the SACT units outputs and routes the operation based on the results and the counter state.
\blue{Figure~\ref{fig:robocore_ctrl} illustrates an example of the control logic for the Box-Normal unit, showing routing for four different queries over time.}
If the output is a separating axis, the controller terminates the program (\blue{i.e. Query 0}), sending the results back to the traversal controller \blue{and updating the warp buffer status}.
Otherwise, the controller returns the operation to the same SACT unit for the next axis test (\blue{i.e. Query 2 and 3}), or forwards the operation to the next SACT unit when the counter overflows (\blue{i.e. Query 6}).
This control mechanism allows for early termination of non-colliding queries and reduces interconnect traffic.
We also include a new RETURN OP unit in the programmable datapath that can be used to signal early exits for non-SACT kernels, which is described in Section~\ref{sssec:return_op}.

\subsection{Early Exit Support for Non-SACT Kernels}
\begin{figure}[t]
    \centering
    \includegraphics[width=0.5\textwidth]{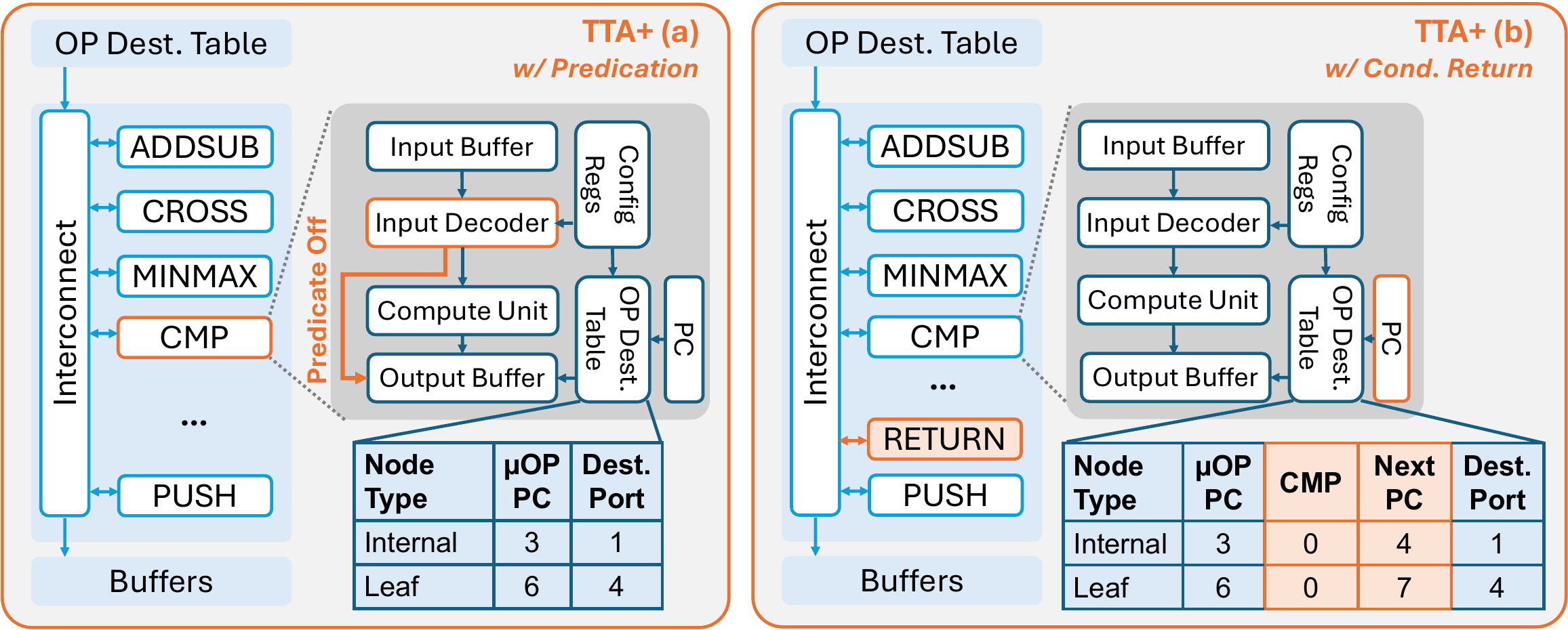}
    \caption{Early exit support on TTA$+$ with (a) predication and (b) conditional returns. Modifications are in orange.}
    \label{fig:tta_plus}
\end{figure}

The \roboRT datapath adds early exits for the SACT kernel, but early exits are beneficial in other robotics kernels as well.
Figure~\ref{fig:tta_plus} illustrates two possible approaches to support non-SACT $\mu$op programs generically applicable to TTA$+$ architecture: predication and conditional returns.

\subsubsection{Predication}
Predication requires minimal hardware modifications by tagging each $\mu$op with a predicate flag stored as part of the interconnect packet.
The input decoder uses this bit to control whether the operation is executed or bypassed directly to the output buffer.
The CMP (compare) unit updates the predicate based on comparison outcomes, similar to control tokens in dataflow machines~\cite{dennis1974preliminary}.
Although predication can suppress arithmetic work, all $\mu$ops still need to be decoded and travel through the interconnect, which still incurs an overhead and wastes energy.
Section~\ref{ssec:eval_earlyexit} evaluates the performance and inefficiencies of predication.

\subsubsection{Conditional Returns}
\label{sssec:return_op}
A better approach is using conditional returns, where CMP redirects control flow like a switch operator in dataflow machines~\cite{arvind1987executing} and skips the remaining program.
To support this behavior, we add a RETURN OP unit (Figure~\ref{fig:tta_plus} (b)) that terminates the intersection program and signals back to the traversal controller that traversal can continue. 
The same RETURN OP unit is included in \roboRT.
Each OP unit in TTA$+$ relies on an OP destination table to specify where the output data should be routed next, but the RETURN OP unit does not need one since it always signals the end of the program.
In the CMP OP unit, we extend the OP destination table with a column for the compare result to create two entries per $\mu$op PC for true and false outcomes that route to different next OP units.
The PC counter inside CMP is modified to read the next PC value from the OP destination table instead of simply incrementing.
These changes support conditional branching in general beyond just returns, which opens up possibilities for more complex intersection programs in the future.
Section~\ref{ssec:eval_earlyexit} compares the performance of conditional returns versus predication.

\subsection{\blue{Generalizing \roboRT}}
\label{sec:pointcloud}

\begin{figure}[t]
\centering
\includegraphics[width=0.49\textwidth]{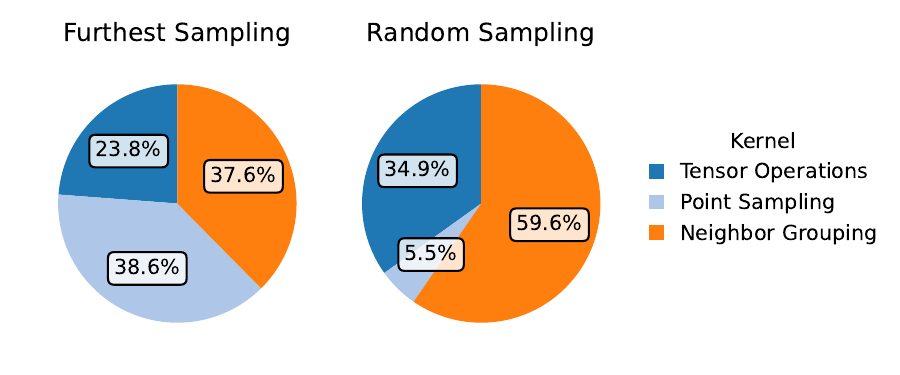}
\caption{Latency distribution of M$\pi$Net~\cite{mpinet} using furthest point sampling and modified random point sampling.}
\label{fig:latency}
\end{figure}

\blue{Other robotics algorithms with similar GPU inefficiencies can also be mapped to the \roboRT functionality.}
\blue{\roboRT supports any kernel that can execute on RTAs or TTA$+$, which already includes a wide range of tree traversal and pointer-chasing algorithms.}
One important \blue{example} non-SACT kernel that \roboRT accelerates is used in point cloud processing, which motion planners rely on to understand the environment.
Point clouds are obtained from sensors such as LiDAR then processed through networks like PointNet++~\cite{pointnetplus} to extract features.
Figure~\ref{fig:latency} shows the latency distribution of a single  M$\pi$Net~\cite{mpinet} forward pass executed on a GPU, which uses PointNet++ as the backbone.
The neighbor grouping kernel is the biggest bottleneck overall, consuming 59.6\% of the total inference time for random sampling and 37.6\% for furthest point sampling.
This kernel is executed as a ball query operation that identifies neighboring points within a certain radius (ball) of the query, which can already be reformulated to RTAs as described in RTNN~\cite{zhu2022rtnn} and FastRNN~\cite{RT_RadiusSearch} using spheres and rays. 
Although the reformulated RTA implementation performs better than CUDA, custom intersection programs for ray-sphere tests and the lack of early-exit support incur high overheads on RTAs.
Fortunately, both issues are resolved by \roboRT, meeting our goal of an accelerator that adapts to different robotics kernels.

\begin{figure}[t]
    \centering
    \includegraphics[width=0.45\textwidth]{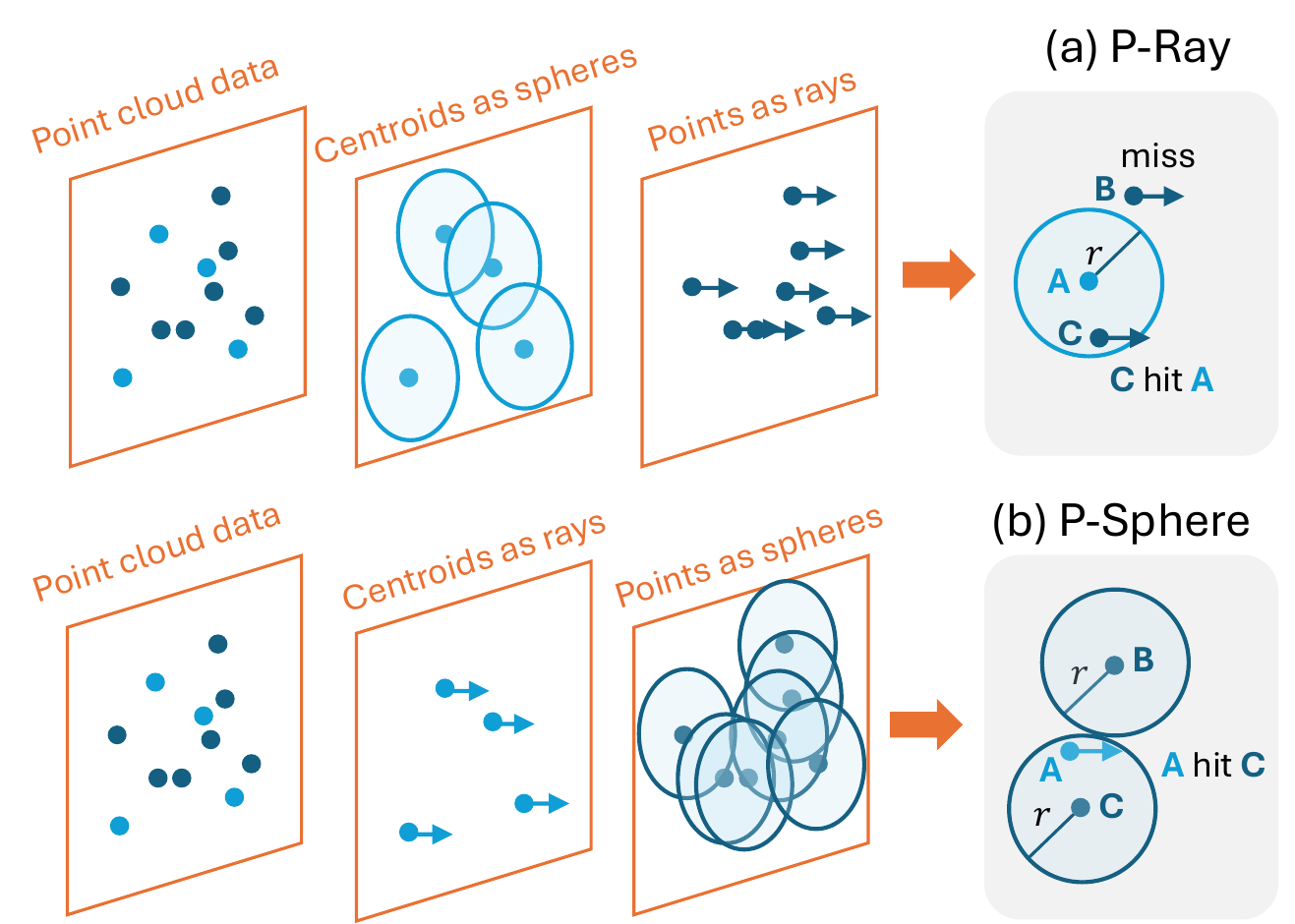}
    \caption{Approaches for applying ray tracing to PointNet++. (a) \rtnnConfigA: sampled points as spheres and others as rays. (b) \rtnnConfigB: sampled points as rays and others as spheres.}
    \label{fig:rtnn}
\end{figure}

There are two different ways to execute ball query on \roboRT since the operation is commutative~\cite{RT_RadiusSearch}. 
Figure~\ref{fig:rtnn} illustrates the two approaches (\rtnnConfigA and \rtnnConfigB), where \rtnnConfigA models the sampled points as spheres and other points as rays, and \rtnnConfigB models the sampled points as rays and other points as spheres.
For example, if point A is the sampled point, and the goal is to find if point B or point C is within the radius of point A, then \rtnnConfigA creates a sphere around point A and traces infinitely short rays from point B and point C to check for intersections, while \rtnnConfigB creates spheres around points B and C and traces infinitely short rays from point A to check for intersections.
In both cases, the result is that ray A hits sphere C or ray C hits sphere A, implying that point C is within the desired radius of point A.
Although FastRNN introduces both configurations, the authors only evaluate \rtnnConfigA based on the assumption that RTAs favor \rtnnConfigA~\cite{RT_RadiusSearch}.
With the added support in \roboRT, we expect \rtnnConfigB surpasses \rtnnConfigA because more spheres are built into highly efficient bounding volume hierarchies (BVH) structures without additional ray-sphere test overheads and can be skipped using early exits once enough points are identified.
We evaluate both configurations in Section~\ref{sssec:pointcloud_processing}.

\subsection{\blue{\roboRT Programming Model}}
\blue{
Programming \roboRT is similar to programming existing RTAs, in which intersection programs are pre-configured to the device then invoked during tree traversal.
The \roboRT datapath can be configured to execute either the default SACT program or custom $\mu$op sequences that allow developers to implement new kernels as needed.
Table~\ref{table:api} summarizes the API for programming \roboRT and Table~\ref{table:robocore-ops} lists the available $\mu$ops with their associated OP unit and description.
}
\blue{
\roboRT functions are invoked within a GPU kernel where each thread represents a single collision query.
The programmer needs to configure \roboRT to decode the data layout of the OBBs and octree nodes using the \texttt{ConfigDecode} API.
Then, each thread executes \texttt{CheckCollisions}, which launches the SACT program on \roboRT and returns the collision results after completion.
Listing~\ref{list:apiexample} summarizes the API usage with pseudocode for collision detection on \roboRT, which directly integrates with CUDA kernels for general-purpose programming, unlike TTA$+$ that uses the Vulkan graphics pipeline.
For other programs, the host first configures the desired datapath on \roboRT by sending a $\mu$op assembly file that specifies the sequence of operations and their configurations, much like the TTA$+$ programming model.
}

\blue{
Another benefit of the \roboRT programming model is the ability to track tree traversal metrics such as the number of nodes visited.
This feature is enabled by decoupling \roboRT from the graphics paradigm and allowing additional variables to be pushed to the warp buffer through the PUSH $\mu$op.
We demonstrate how tracking these metrics can be used to further optimize traversal performance by adjusting traversal strategies dynamically based on program characteristics at runtime in Section~\ref{sssec:rowild}.
}

\begin{table}[t]
    \caption{\blue{\roboRT Programming API}}
    \label{table:api}
    \centering
    \scriptsize
    \begin{tabularx}{0.48\textwidth}{|l|X|}
    \hline
    \textbf{Type} & \textbf{API / Description} \\ \hline
    Host & \texttt{ConfigRoboCore("collision\_alg.asm")} \\
    & Configure desired datapath on \roboRT using $\mu$op assembly file. \\ \hline
    Host & \texttt{ConfigDecode(Query, NodeA, NodeB)} \\
    & Configure decode units on \roboRT for query, node type A (e.g. inner nodes), and node type B (e.g. leaf nodes) layouts. \\ \hline
    Device & \texttt{CheckCollisions(OBB, TreeRoot, Results)} \\
    & Defaults to SACT algorithm. Executes collision detection on the input OBB and octree. \\ \hline
    Device & \texttt{ExecuteRoboCore(Query, TreeRoot, Results)} \\
    & Runs configured datapath (e.g. ball query) on the input query and tree structure. \\ \hline
    \end{tabularx}%
\end{table}

\begin{table}[t]
    \caption{\blue{Supported \roboRT $\mu$ops}}
    \label{table:robocore-ops}
    \centering
    \scriptsize
    \begin{tabularx}{0.48\textwidth}{|l|l|X|l|}
    \hline
    \textbf{OP Unit} & \textbf{Supported $\mu$op} & \textbf{Description} & \textbf{Latency} \\ \hline
    ADDSUB & ADD, SUB & FP32 Vec3 addition and subtraction. & 3 cycles\\ \hline
    MINMAX & MINMAX & FP32 MIN(a, MAX(b, c)). & 2 cycles\\ \hline
    MAXMIN & MAXMIN & FP32 MAX(a, MIN(b, c)). & 2 cycles\\ \hline
    CMP & CMP & Per-element Vec3 comparison. & 1 cycle\\ \hline
    MUL & MUL & FP32 scalar multiplication. & 5 cycles\\ \hline
    Pre-Process & PRE, XFORM & SACT preprocessing operations, ray transform matrix multiplication. & 14 cycles\\ \hline
    Box-Normal & BOX, DOT & SACT Box-Normal operation, FP32 Vec3 dot product. & 12 cycles\\ \hline
    Edge$\times$Edge & EDGE, CROSS & SACT Edge$\times$Edge operation, FP32 Vec3 cross product. & 12 cycles\\ \hline
    LOGIC & AND, OR, XOR, NOT & Logical operations. & 1 cycle\\ \hline
    PUSH & PUSH & Push to warp buffer or traversal stack. & 1 cycle\\ \hline
    RETURN & RETURN & Exit from OP units. & 1 cycle\\ \hline
    \end{tabularx}%
\end{table}

\begin{lstlisting}[
    belowskip=-2\baselineskip,
    backgroundcolor=\color{backcolour},   
    commentstyle=\color{codegreen},
    numberstyle=\tiny\color{codegray},
    stringstyle=\color{codegray},
    basicstyle=\tiny,
    breaklines=true,
    captionpos=b,
    keepspaces=true,
    numbers=left,
    xleftmargin=2em,
    numbersep=5pt,
    float=t, 
    language=C++,
    showstringspaces=false,
    caption={\blue{Programming \roboRT for collision detection.}}, 
    label={list:apiexample}]
// Define OBBs and octree structures for robot poses and the environment
OctreeNode* treeRoot = BuildOctree(EnvironmentPointCloud);
OBB* obbs = CreateOBBs(RobotPoses);
// Copy to device and allocate results buffer
CollisionResult* d_results = AllocateDeviceBuffer();
OBB* d_obbs = CopyToDevice(obbs);
OctreeNode* d_treeRoot = CopyToDevice(treeRoot);

// Configure OBB and octree node layouts (byte offsets)
size_t obbLayout[5] = {0, 6, 12, 18, 24};
size_t nodeLayout[3] = {0, 2, 10};
ConfigDecode(obbLayout, nodeLayout, nodeLayout);

// Define kernel for collision detection
__global__ void CollisionKernel(OBB* d_obbs, OctreeNode* d_treeRoot, CollisionResult* d_results) {
    int idx = blockIdx.x * blockDim.x + threadIdx.x;
    CheckCollisions(d_obbs[idx], d_treeRoot, &d_results[idx]);
}

// Launch kernel with appropriate dimensions and synchronize results
CollisionKernel<<<blocks, threadsPerBlock>>>(d_obbs, d_treeRoot, d_results);
cudaDeviceSynchronize();
CollisionResult* h_results = CopyToHost(d_results);
\end{lstlisting}

\section{Methodology}
\label{sec:methodology}

We evaluate our proposed \roboRT architecture using Vulkan-Sim~\cite{vulkansim}, a cycle-level GPU architecture simulator \blue{commonly used to evaluate ray tracing hardware~\cite{ha2024tta,chou2025treelet,tozlu2025cooprt,huang2025aqb8,feng2025heliostat,saed2025rayn,wang2025process}}.
\blue{We rely on Vulkan-Sim to measure \roboRT performance since \roboRT is not in any existing robots and robotics simulators such as Ros{\'e}~\cite{nikiforov2023rose} are not designed to evaluate GPU systems.}
Table~\ref{table:simconfig} shows the hardware configurations used in our evaluation, matching TTA$+$.
We model \roboRT with SACT units in Vulkan-Sim to include our architectural changes.
We evaluate energy consumption of \roboRT using AccelWattch~\cite{kandiah2021accelwattch} integrated with Vulkan-Sim and the OP units in \roboRT by tracking active cycles and operations performed by each unit.
To estimate the area overhead, we implement all OP units using Chisel~\cite{bachrach2012chisel}, with Berkeley HardFloat~\cite{hauser_hardfloat_2019} for floating point operations, and synthesize with Synopsys Design Compiler using the FreePDK 45nm standard cell library~\cite{stine2007freepdk}.
The synthesized design is used to estimate latency for each SACT unit in the simulator. 

\begin{table}[t]
    \caption{Vulkan-Sim Configurations}
    \label{table:simconfig}
    \resizebox{\columnwidth}{!}{%
    \centering
    \scriptsize
    \begin{tabular}{|l|cc|}
    \hline
    \# Streaming Multiprocessors (SM) & \multicolumn{2}{c|}{8}                             \\ \hline
    Max Warps / SM                    & \multicolumn{2}{c|}{32}                                \\ \hline
    Warp Scheduler                    & \multicolumn{2}{c|}{GTO}                               \\ \hline
    \# Registers / SM                 & \multicolumn{2}{c|}{32768}        \\ \hline
    Instruction Cache                 & \multicolumn{2}{c|}{128KB, 16-way assoc., 20 cycles}   \\ \hline
    L1 Data Cache + Shared Memory     & \multicolumn{2}{c|}{64KB, Fully assoc. LRU, 20 cycles} \\ \hline
    L2 Unified Cache                  & \multicolumn{2}{c|}{3MB, 16-way assoc. LRU, 160 cycles} \\ \hline
    Compute : Interconnect : L2 : Memory Clock
                                      & \multicolumn{2}{c|}{1365 : 1365 : 1365 : 3500 MHz}                          \\ \hline
    \# \roboRT Units / SM                  & \multicolumn{2}{c|}{1}                                 \\ \hline
    Warp Buffer Size                  & \multicolumn{2}{c|}{4 warps}                                 \\ \hline
    \# \roboRT Intersection Units         & \multicolumn{2}{c|}{4 sets}                                 \\ \hline
    \end{tabular}%
    }
\end{table}

\subsection{Evaluation Workloads}
We evaluate the performance of \roboRT across several robotics workloads, each using their associated environment datasets, and compare against baselines described in this section.

\subsubsection{Environments}
We evaluate collision detection on four different environments with varying complexity and density of obstacles: Cubby, Dresser, Merged Cubby, and Tabletop.
These environments are taken from M$\pi$Net~\cite{mpinet} and each environment is paired with OBBs enclosing different robot poses along planned trajectories.
The Tabletop environment is the simplest structure, but still features randomly placed obstacles that create a cluttered environment.
The Cubby, Dresser, and Merged Cubby environments have a more complex structure, with smaller openings that require precise motion planning.
To compare against MPAccel~\cite{energyefficientmotionplanning2024shah}, we additionally evaluate ten environment scenarios with 100 pairs of start and end goals per environment based on data published by the authors.
We also evaluate performance with Alpamayo-R1~\cite{alpamayo} as a representative VLA model, using the public 10B parameter model with randomly selected clips from Physical AI AV Dataset~\cite{nvidia_physicalai_av_2025}, which provides LiDAR data that we use for collision detection.
These environments are static during the collision query and we leave dynamic obstacles to future work. 

\subsubsection{Collision Detection}
M$\pi$Net does not include collision detection in their code even though the authors acknowledge that it should be combined for safe operation in real robot systems, so we implement our own OBB-octree collision queries. 
We use the same programming model as TTA$+$, with each thread representing a single OBB-octree collision query.
We compare the performance of \roboRT against a CUDA-based GPU implementation~\cite{energyefficientmotionplanning2024shah} that also checks for collisions by testing OBBs against an octree data structure as our baseline.
We also compare against Mochi~\cite{mandarapu2024mochi}, which maps collision detection to RTA hardware on GPUs.
However, Mochi is designed for particle-based simulations that track all collisions, whereas motion planning applications only need to detect collisions between the robot and the environment.
Therefore, we modify Mochi to filter out self-collisions using the RTA programmable intersection shaders.
We build a BVH tree using point cloud data of the environment and robot OBBs with Embree~\cite{wald2014embree}, implement Mochi using the Vulkan API~\cite{vulkan}, and evaluate its performance using Vulkan-Sim.
\blue{
Lastly, we compare against CuRobo~\cite{curobo}, an optimized CUDA robot library that includes collision detection.
However, CuRobo employs spheres instead of OBBs for its robot representation even though it is less accurate and results in more false positive collisions.
We implement a sphere-based collision detection on \roboRT using our programmable $\mu$ops to compare against CuRobo as a reference. 
}

\subsubsection{Overall Neural Planning}
We evaluate the performance of M$\pi$Net~\cite{mpinet} with acceleration from \roboRT on the ball query kernel in PointNet++~\cite{pointnetplus}, comparing to the RTNN implementation provided by the authors~\cite{zhu2022rtnn}.
We use the same PointNet++ configurations as implemented in M$\pi$Net to ensure a fair comparison.
For measuring the latency of other kernels that do not require hardware modifications, we use the CUDA implementation provided by the authors of M$\pi$Net~\cite{mpinet}, measured on a NVIDIA RTX 5070 Ti GPU.
We also evaluate \roboRT with MPNet~\cite{mpnet} despite its lower success rates to create a direct performance comparison against other hardware platforms, including MPAccel~\cite{energyefficientmotionplanning2024shah} (standalone accelerator) and P3NetCore~\cite{sugiura2024integrated} (FPGA). 
To our knowledge, there are no other hardware solutions that target neural planners with explicit collision detection.

\subsubsection{Classical Motion Planning}
Although \roboRT is designed for collision detection and neural motion planning, we also evaluate its performance on a classical algorithm from the RoWild benchmark~\cite{rowild}.
RoWild includes a set of environments and tasks commonly encountered in robotics applications, such as navigation, manipulation, and perception.
While most kernels in RoWild are already efficiently parallelized for GPUs, we find that some irregular kernels can benefit from \roboRT.
For example, DeliBot uses a ray casting algorithm for Monte Carlo Localization that accounts for 74\% of end-to-end latency.
We modify the localization kernel to use the \roboRT architecture for ray casting, which traverses by stepping along the ray rather than down a tree and checks for collisions by accessing the occupancy grid of the environment.

\section{Results}
\label{sec:results}
We find that a RoboGPU with \roboRT effectively accelerates collision detection and overall neural planning compared to a baseline GPU system with an RTA.
In this section, we present our evaluation results for \roboRT.

\subsection{Collision Detection Performance}
\label{ssec:eval_earlyexit}

\begin{figure}[t]
    \centering
    \includegraphics[width=0.5\textwidth]{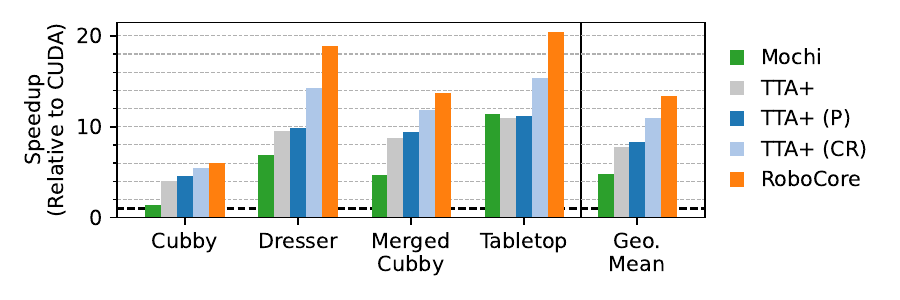}
    \caption{Speedup of collision detection on \roboRT relative to CUDA baseline, with comparisons to Mochi on RTAs, baseline TTA$+$, and TTA$+$ modifications for predication (P) and conditional returns (CR).}
    \label{fig:collision_detection_results}
\end{figure}

\begin{figure}[t]
    \centering
    \includegraphics[width=0.5\textwidth]{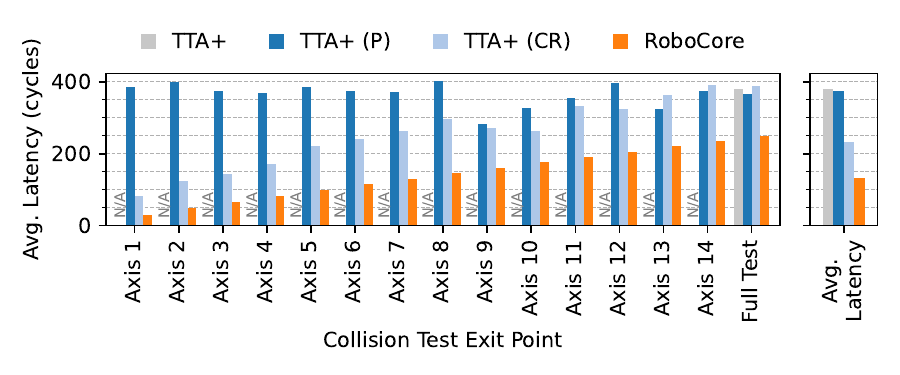}
    \caption{Latency distribution of collision detection queries for different exit conditions. TTA$+$ only performs Full Test.}
    \label{fig:collision_latency}
\end{figure}

\roboRT achieves \speedupCollisionCuda and \speedupCollision speedup on collision detection queries over the CUDA and Mochi baselines, respectively, as shown in Figure~\ref{fig:collision_detection_results}.
Although Mochi already accelerates collision detection using RTAs, its performance suffers from many unnecessary collision checks and the cost of invoking the intersection shader at each hit.
In contrast, the \roboRT architecture is optimized for collision detection and avoids overheads that are created by forcing queries through the ray tracing pipeline.
Early termination and collision units within \roboRT also leads to significant improvements over the TTA$+$ architecture.
Figure~\ref{fig:collision_latency} shows the latency distribution of collision queries for different exit conditions, highlighting the benefits of early termination.
Predication is an ineffective early termination option since $\mu$ops must still travel through the interconnect and OP units, making almost no impact on the average collision latency.
Conditional returns (CR) eliminate this overhead, providing 40\% speedup over TTA$+$ alone.
Finally, dedicated SACT units on \roboRT create an additional 22\% speedup over TTA$+$ (CR), benefitting from specialization while preserving flexibility.

\begin{figure}[t]
    \centering
    \includegraphics[width=0.5\textwidth]{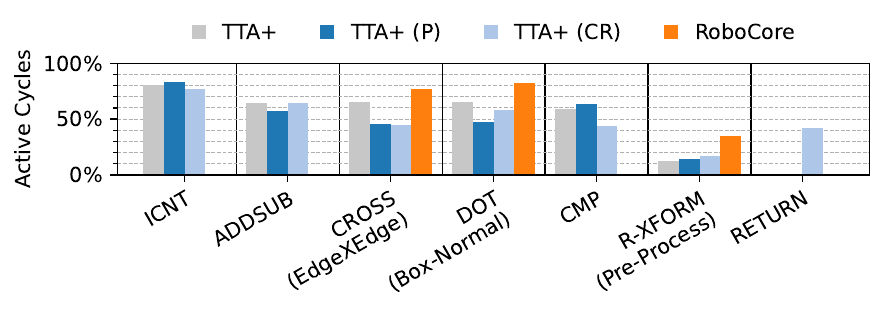}
    \caption{Utilization of \roboRT OP units for each implementation. Unused OP units are omitted.}
    \label{fig:cdu_util}
\end{figure}

\begin{figure}[t]
    \centering
    \includegraphics[width=0.5\textwidth]{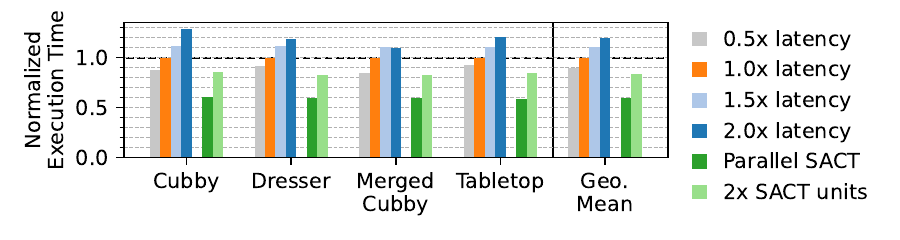}
    \caption{\blue{Relative impact of SACT unit configuration on overall collision detection performance.}}
    \label{fig:latency_sweep}
\end{figure}

These advantages can be observed in Figure~\ref{fig:cdu_util}, which shows the utilization of each \roboRT OP units for SACT.
TTA$+$ configurations show very high interconnect (ICNT) utilization caused by unnecessary data movements, adding latency and costing energy.
With \roboRT, only the SACT units are needed for collision detection, leading to much better utilization within the specific units and the possibility to reduce energy by power gating all other OP units.
Furthermore, Figure~\ref{fig:latency_sweep} compares the performance impact of different SACT unit latency, adjusting by 0.5$\times$ to 2$\times$ on all SACT units.
The normalized execution time of collision detection shows only small performance variations, implying insensitivity to individual SACT unit latency and suggesting that the SACT units can be designed to conserve area and energy instead of optimizing for latency despite collision detection being a time-sensitive task.
\blue{
Figure~\ref{fig:latency_sweep} also compares \roboRT to a parallel SACT design, where six Box-Normal and nine Edge$\times$Edge units are used to perform all 15 axis tests simultaneously.
While latency is reduced by 41\%, the area is nearly 3.7$\times$ larger.
If latency is prioritized over hardware area, a better design point could include two of each SACT unit in \roboRT, which results in $1.2\times$ speedup at $1.8\times$ the area of the default option and maintains early exit energy savings.
}

\begin{figure}[t]
    \centering
    \includegraphics[width=0.5\textwidth]{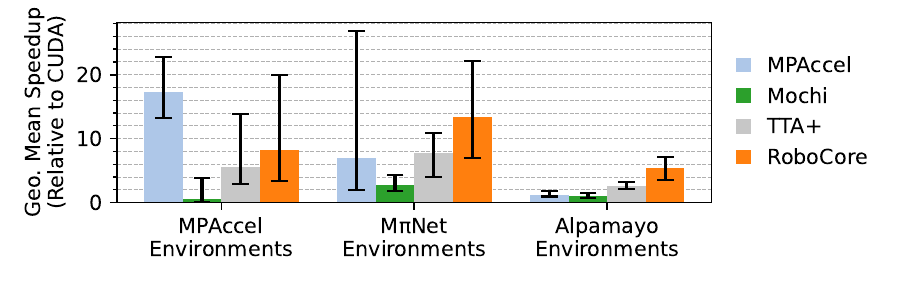}
    \caption{Performance results of collision detection for MPAccel test scenarios compared to M$\pi$Net and Alpamayo-R1 environments, showing average, min, and max speedup.}
    \label{fig:mpaccel_comparison}
\end{figure}

\roboRT can also accelerate other scenarios, such as smaller MPAccel environments~\cite{energyefficientmotionplanning2024shah} and Physical AI AV scenes~\cite{nvidia_physicalai_av_2025} with simpler robot geometries, supporting a range of real-world behavior. 
Figure~\ref{fig:mpaccel_comparison} compares \roboRT to MPAccel.
Although \roboRT cannot match MPAccel on their environments, which reports 23$\times$ speedup using 16 collision units over the same CUDA baseline for $2^{20}$ collision queries, \roboRT benefits from a more sophisticated memory system and eliminates IO overheads that standalone accelerators incur.
Figure~\ref{fig:mpaccel_comparison} also evaluates \roboRT on randomly selected scenes from Physical AI AV~\cite{nvidia_physicalai_av_2025} used by Alpamayo-R1~\cite{alpamayo}.
Vehicles have fewer degrees of freedom than the robotic arms in M$\pi$Net but \roboRT is still 5.4$\times$ faster on average, generalizing better to different use cases than MPAccel.

\blue{
Using sphere-based robot representation, \roboRT achieves 21$\times$ speedup over the OBB-based CUDA baseline and is faster than CuRobo~\cite{curobo}.
CuRobo latency ranges from 4.6-9.9$\mu$s per robot pose, compared to 0.18-0.36$\mu$s on \roboRT for the same M$\pi$Net environments.
However, CuRobo does not use an octree structure and computes a complex distance metric in collision detection, accounting for its higher latency.
Regardless, we find that sphere approximations results in 10\% false collisions, causing some feasible trajectories to be rejected.
Unnecessary replanning is then required, leading to worse overall performance for the motion planner.
}

\begin{figure}[t]
    \centering
    \includegraphics[width=0.5\textwidth]{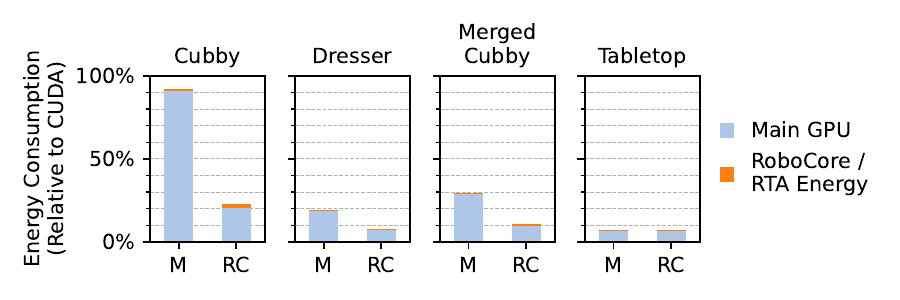}
    \caption{Energy consumption of collision tests for Mochi (M) and \roboRT (RC).}
    \label{fig:energy}
\end{figure}

The performance improvements of \roboRT directly translates to energy savings, which is a key factor in robotics.
Figure~\ref{fig:energy} shows the energy consumption results of \roboRT for collision detection queries compared to the CUDA and Mochi baselines.
Energy from the main GPU components dominate the total energy consumption, with the \roboRT only contributing to less than 5\% of the total energy on average.
We achieve 87\% energy savings against the CUDA baseline and \energysavings against Mochi by reducing redundant operations and improving overall performance.

\subsection{Neural Planning Performance}
\roboRT outperforms existing point cloud processing implementations such as RTNN, which compounds with the fast collision detection to accelerate end-to-end robotics motion planning.

\subsubsection{Point Cloud Processing}
\label{sssec:pointcloud_processing}

\begin{figure}[t]
    \centering
    \includegraphics[width=0.5\textwidth]{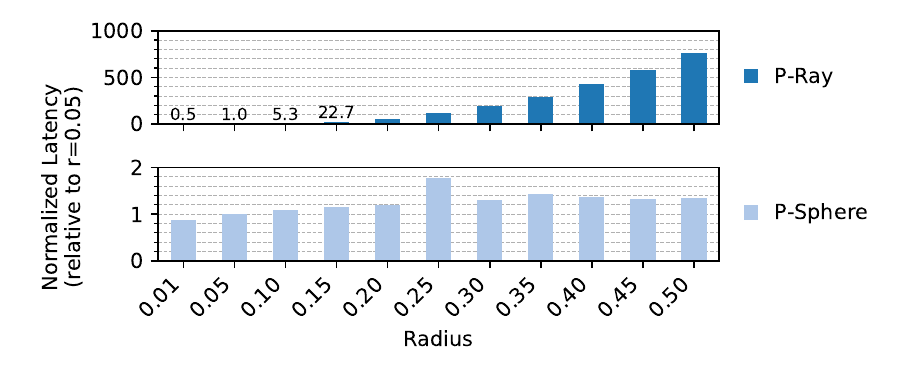}
    \caption{Execution latency of different grouping radii (r) on ball query with \roboRT, relative to r=0.05 used by M$\pi$Net. }
    \label{fig:rtnn_radius}
\end{figure}

\begin{table}[t]
    \caption{Comparison of \rtnnConfigA and \rtnnConfigB on PointNet++ for Cubby environment with \roboRT relative to RTNN.}
    \label{table:rtnn_comparison}
    \centering
    \begin{tabular}{|l|c|c|}
    \hline
    \textbf{} & \textbf{\rtnnConfigA}  & \textbf{\rtnnConfigB} \\ \hline
    \textbf{Total rays} & 523,776 & 512 \\ \hline
    \textbf{\roboRT occupancy} & 100\% & 47.4\% \\ \hline
    \textbf{Total spheres} & 512 & 523,776 \\ \hline
    \textbf{Tree depth} & 6 & 12 \\ \hline
    \textbf{Avg. (nodes, spheres) / ray} & (14.1, 2.3) & (959.2, 276.1) \\ \hline
    \textbf{Total nodes traversed} & 6,856,980 & 349,272 \\ \hline
    \textbf{Speedup} & 1.19$\times$ & 2.93$\times$ \\ \hline
    \hline
    \textbf{Avg. speedup (all env.)} & 1.17$\times$ & 1.95$\times$ \\ \hline
    \end{tabular}%
\end{table}

We find that the CUDA implementation of ball query used in PointNet++ is inefficient and RTNN provides a more reasonable baseline.
RTNN has been shown to reach 44$\times$ speedup on average even over the highly optimized cuNSearch~\cite{hoetzlein2014fast} implementation.
\roboRT further outperforms RTNN by 1.9$\times$ on average using \rtnnConfigB, which is only made possible by our proposed architectural optimizations.
\rtnnConfigB is enhanced by the programmability and early termination support in \roboRT, which avoids the overhead of running custom intersection shaders in the compute cores and unnecessary intersection tests beyond the required number of spheres.
Table~\ref{table:rtnn_comparison} compares \rtnnConfigA and \rtnnConfigB for the Cubby environment to quantify this difference.
\rtnnConfigA generates a large number of rays that traverse through a small tree of spheres with six levels, whereas \rtnnConfigB generates a small number of rays that traverse through a much larger tree of spheres with twelve levels.
Surprisingly, \rtnnConfigB performs better despite only launching enough rays to utilize 47.4\% of the \roboRT occupancy.
This result is because the tree structure organizes the spheres for a very efficient traversal, leading to fewer memory fetches for node data overall.
In the RTNN implementation, there is no mechanism to immediately stop traversal after a maximum group size is reached, forcing the hardware to continue traversing the tree and launching shader invocations simply to find that no more points are needed.
This problem is much worse with \rtnnConfigB because the number of spheres is much larger than the maximum group size.
On average, \roboRT traverses 6$\times$ fewer nodes with early exit than without, which leads to a significant reduction in latency and energy consumption.

To ensure these results generalize beyond M$\pi$Net, we also sweep different radius values that may be used by PointNet++ in other robotics workloads like ContactGen~\cite{liu2023contactgen}, Sugar~\cite{chen2024sugar}, HDP~\cite{ma2024hierarchical} or others.
Figure~\ref{fig:rtnn_radius} shows that \rtnnConfigA does not scale well with larger radius values and the relative latency grows exponentially alongside the increased intersection shader invocations.
\rtnnConfigB, which is enabled by \roboRT, scales much better.

\subsubsection{Full Motion Planning Pipeline}
\label{ssec:full_pipeline}

\begin{figure}[t]
    \centering
    \includegraphics[width=0.5\textwidth]{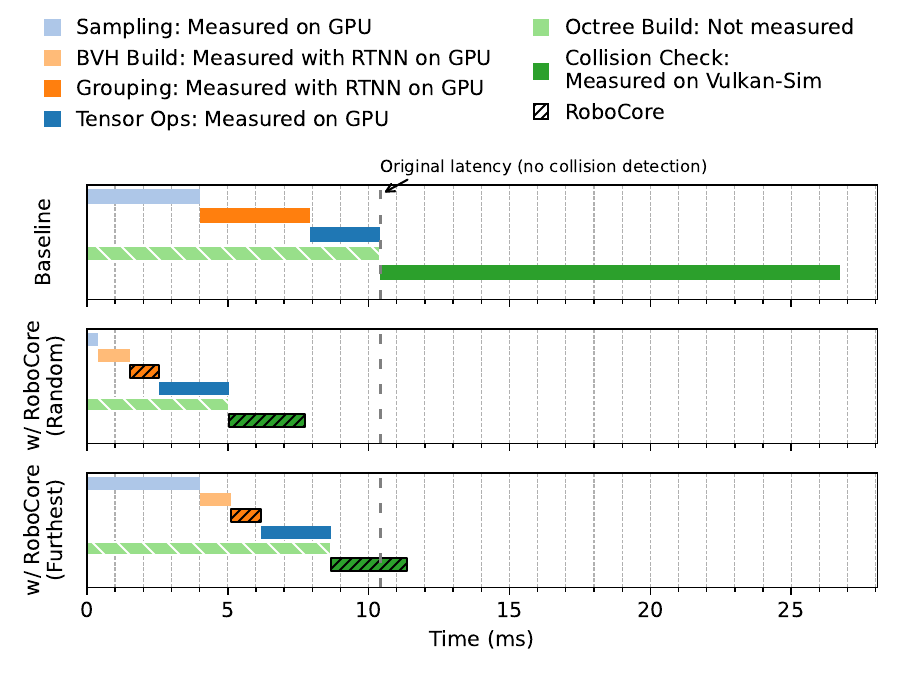}
    \caption{Latency breakdown of proposed motion planning pipeline on GPU with \roboRT compared to the baseline CUDA implementation for Cubby environment.}
    \label{fig:e2e}
\end{figure}

By accelerating point cloud processing in PointNet++, we balance neural inference and \roboRT tasks, allowing better utilization of the specialized GPU resources in an end-to-end motion planning pipeline.
Figure~\ref{fig:e2e} shows the latency breakdown of M$\pi$Net measured on a GPU, including the neural motion planner and explicit collision detection. 
Since \roboRT cannot be measured on existing GPUs, we estimate the BVH Build and Grouping times using RTNN and the Collision Check time by applying our measured speedup over CUDA.
Using random sampling in PointNet++, combined with neighbor grouping on \roboRT, we accelerate the point cloud processing stage in M$\pi$Net by 2.2$\times$ over the CUDA baseline.
With this speedup, the additional latency from explicit collision detection does not delay the overall motion planning pipeline and actually completes 1.3$\times$ faster on our RoboGPU with \roboRT than the baseline \blue{while also reducing the 14\% collision rate to 0\%}.
Although PointNet++ uses furthest point sampling by default, we find that random point sampling can achieve an \randomSuccessRate success rate, which is only slightly lower than the \furthestSuccessRate of furthest point sampling, but saves 29\% of the total latency.
\blue{This result is important as it has been shown that lower latency can be more important than higher accuracy for overall robotics performance~\cite{nikiforov2023rose}}.
The full pipeline is \speedupMpiNet faster and meets real-time latency requirements for robotics, overcoming the limitations described in M$\pi$Net for enabling safer operations.

Octree construction can begin as soon as the point cloud is available, in parallel with other motion planning tasks. 
The total latency is dependent on the speed of the octree builder, which \blue{only becomes the critical path if it is slower than the sensor frequency when combined with collision checking}.
However, there are many orthogonal works optimizing octree construction and updates~\cite{deng2017g, chen2025octocache, min2023octomap,durvasula2022voxelcache} that can be integrated, \blue{which is beyond the scope of \roboRT}.
After the first iteration octree build, subsequent octree updates are substantially faster \blue{and can update at around 20Hz~\cite{chen2025octocache} compared to the 10Hz LiDAR sensor~\cite{nvidia_physicalai_av_2025,xu2026wod}}.
Collision detection and milestone generation in M$\pi$Net may also be pipelined together using accelerator-level parallelism between the \roboRT and a \tensorcore to further reduce latency.

To compare end-to-end performance against other hardware designs, Table~\ref{tab:planning_latency} presents the total planning latency of MPNet on different devices~\cite{energyefficientmotionplanning2024shah,sugiura2024integrated} against a GPU with \roboRT.
\roboRT is faster than both baseline GPUs and an FPGA accelerator (P3NetCore), highlighting the benefits of specialization. 
Although MPAccel is faster, it potentially underestimates coordination overheads between the DNN accelerator, CPU controller, and MPAccel that \roboRT does not have.
MPAccel also relies on strategic scheduling that is only beneficial when a collision exists, which becomes less likely as neural planner improve.
Both P3NetCore and MPAccel lack programmability compared to GPUs and \roboRT, executing only MPNet and P3Net, which are very small MLP models with poor success rates compared to newer neural planners.

\begin{table}[t]
\caption{Overall planning latency for MPNet (including collision detection) on different devices.}
\label{tab:planning_latency}
\centering
\begin{tabular}{|l|c|}
\hline
\textbf{Hardware} & \textbf{Latency (ms)} \\ \hline
NVIDIA RTX 3090 & 21 \\ \hline 
NVIDIA Jetson TX2 & 110 \\ \hline 
P3NetCore (FPGA) & 50\\ \hline 
MPAccel (standalone accelerator) & 1.1 \\ \hline 
GPU w/ \roboRT & 7.5 \\ \hline 
\end{tabular}
\end{table}

\subsection{Classical Algorithm}
\label{sssec:rowild}
\begin{figure}[t]
    \centering
    \includegraphics[width=0.5\textwidth]{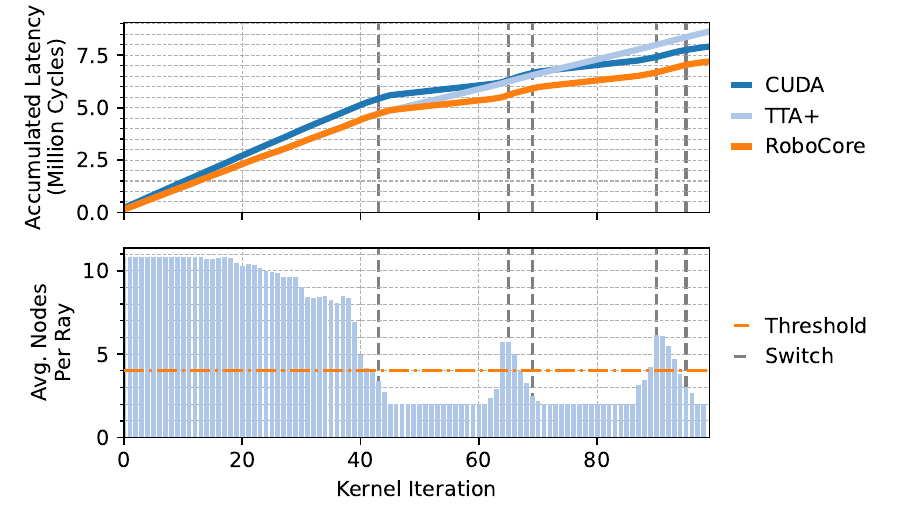}
    \caption{Execution time of DeliBot from RoWild on \roboRT compared to CUDA and TTA$+$ implementations.}
    \label{fig:delibot}
\end{figure}

We demonstrate that \roboRT can improve classical motion planning when neural planners are not suitable using an example from RoWild~\cite{rowild}.
Figure~\ref{fig:delibot} shows the cumulative execution time of the DeliBot kernel on \roboRT compared to CUDA and TTA$+$ implementations, with 10\% and 20\% speedups respectively.
RoWild benchmarks execute over a large number of kernel iterations, exhibiting different characteristics throughout execution. 
For example, DeliBot begins with randomly initialized particles that are refined over time. 
To start, there are few collisions, leading to long ray traversals ideal for \roboRT.
However, as the particles converge, the number of collisions increase, leading to short traversals that fail to offset the \roboRT launch overhead.
Therefore, we allow DeliBot to dynamically switch between execution on the \roboRT and CUDA cores for each kernel iteration, based on expected collisions.
We implement this dynamic switching by using a threshold on average traversal distance in the previous iteration, tracked using the average number of nodes traversed per ray.
Figure~\ref{fig:delibot} also shows how this metric changes over time, triggering the switches between the \roboRT and CUDA cores, resulting in the best overall performance.
This result reinforces the importance of placing \roboRT within the GPU to reap the benefits of both programmability and specialization without incurring orchestration overheads.

Other RoWild kernels like CarriBot and MoveBot can also benefit from \roboRT.
CarriBot uses an A* algorithm that relies on collision detection and MoveBot uses a RRT* algorithm that implements a nearest neighbor search, both of which \roboRT can accelerate.
\blue{
\roboRT can also still be used to accelerate regular ray tracing workloads, with only a 12\% slowdown compared to RTAs when evaluated on LumiBench~\cite{lumibench}. 
This is negligible compared to the $10\times$ speedup achieved by RTAs over software approaches~\cite{rtcore}.
}

\subsection{Area Overheads}
The area overhead of the modifications in the \roboRT architecture is only 17\% more than TTA$+$ since the SACT units are designed to be dual-purpose.
Conditional returns only require an increase to the size of the operation destination table, where each entry is just 2 bytes.
Although we add an additional operation unit for RETURN, the additional port does not increase the area overhead because TTA$+$ already supports up to 16 ports to the interconnect.
Alternatively, the design from MPAccel~\cite{energyefficientmotionplanning2024shah} for collisions only requires an area of 0.251mm$^2$, which is an additional 59\% overhead on top of the TTA$+$ in order to provide the same generality as \roboRT.
Furthermore, MPAccel uses fixed-point hardware unlike \roboRT, which may not be effective for collision detection across a wide variety of environment scales and robot sizes.
Overall, \roboRT is only \areaoverhead larger than an RTA~\cite{shen2025rayflex} synthesized at the same 45nm process, adding less than 1\% area to a robotics GPU such as the NVIDIA Jetson Orin Nano when scaled.
This result fits our goal of minimizing modifications to reduce design and validation costs.

\begin{table}[t]
\caption{Area breakdown of computation in \roboRT units (R) compared to TTA$+$ (T) and MPAccel.}
\label{tab:robocore_area}
\centering
\begin{tabular}{|rl|r|r|}
\hline
& \textbf{Unit} & \textbf{Area (${\mu}m^2$)} & \textbf{Area Overhead} \\
\hline
R & Pre-Process & 223911 & 41054 (+22\%) \\
T &(R-XFORM) & 182857 &   \\
\hline
R & Box-Normal & 78381 & 15112 (+24\%) \\
T & (DOT) & 63269 &   \\
\hline
R & Edge $\times$ Edge & 121284 & 18473 (+18\%) \\
T & (CROSS) & 102811 &   \\
\hline
R,T & Vec3 ADD & 36159 &   \\
R,T & MUL & 15959 &   \\
R,T & Vec3 CMP & 11595 &   \\
R,T & MINMAX & 7120 &   \\
R,T & MAXMIN & 7190 &   \\
\hline
\hline
& \textbf{TTA+ Total} & \textbf{426960} &  \\
& \textbf{RoboCore Total} & \textbf{501599} & \textbf{74639 (+17\%)} \\
& \textbf{MPAccel} & \textbf{251000} & \textbf{251000 (+59\%)} \\ 
\hline
\end{tabular}
\end{table}
\section{Related Work}
\label{sec:relatedwork}
Computer architecture has an important role in robotics. 
This section discusses related works optimizing robotics tasks and tree building techniques for environment representation.

\subsection{Accelerating Robotics Workloads}
Robotics is a broad field with many different tasks spanning scene representation and motion planning. 
While there have been many proposed hardware accelerators for specific robotic workloads~\cite{chen2023parallelnn, energyefficientmotionplanning2024shah, sacks2018robox, feng2020mesorasi, fusion3d, huang2025dadu, ting2025hiper}, these accelerators lack the flexibility to adapt to the rapidly changing landscape of robotics.
Tartan~\cite{tartan} proposes a general-purpose CPU architecture optimized for robotics, but evaluated solely on algorithmic motion planning workloads. 
As the robotics field shifts from traditional algorithmic approaches to neural-based approaches like RT-1~\cite{rt12022arxiv} and GR00T~\cite{bjorck2025gr00t}, GPUs will be a better general-purpose device to support both approaches.

Motion planning and collision detection in robotics are well-studied problems and there have been many works proposing optimizations for these tasks.
Amato and Dale~\cite{770055} first show that motion planning can be parallelized on a multicore CPU, then Gayle et al.~\cite{Gayle} show that both path planning and collision detection can be parallelized on GPUs.
Bialkowski et al.~\cite{parallelrrt} used a brute force CUDA algorithm to parallelize collision detection checks on the GPU, achieving massive speedups over the serial CPU implementation.
Kider et al.~\cite{5509470} showed how a random search algorithm can be accelerated on a GPU.
Pan and Manocha~\cite{doi:10.1177/0278364911429335} proposed a BVH tree-based GPU collision detection using hierarchical traversal, which is faster than prior GPU implementations.
More recently, ReCA~\cite{wan2025reca}, Orianna~\cite{hao2024orianna}, MOPED~\cite{huang2024moped}, and Blitzcrank~\cite{hao2023blitzcrank} all target a specific motion planning algorithm, but are less flexible than a GPU, especially for neural motion planning.
FPGAs~\cite{lienen2020reconros, Atay2006, murray2016microarchitecture, Shi2018, zhang2025sparsity, sugiura2024integrated} and ASICs~\cite{Murray, sorin, Lian2018, daducd, racod} have also been used to accelerate motion planning.
However, FPGAs become prohibitively expensive match GPU levels of neural motion planning support with large power gaps to ASICs~\cite{kuon2006measuring}, while ASICs fail to adapt to the wide variety and changing landscape of motion planning algorithms.
These solutions are less accessible to the consumer market compared to GPUs, but can be more useful in specialized industrial applications.

Typically in robotics, many different algorithms are running concurrently on the same processor. 
Although scheduling is outside the scope of this paper, we note that an efficient scheduling algorithm tailored for the modern robotic workloads is crucial for achieving high performance, such as RED~\cite{li2023red}.
Meanwhile, works such as CoopRT~\cite{tozlu2025cooprt} and treelets~\cite{treeletprefetch, chou2025treelet} introduce 
scheduling optimizations to improve the performance of ray tracing workloads on RTAs, which are likely to be beneficial for \roboRT with robotics workloads as well.
Other solutions also optimize neural inferences to balance workloads between accelerators on the GPU in real-time rendering and can be applied to robotics workloads~\cite{liu2025reframe}.

\subsection{Tree Building Techniques}
While tree building is not in the scope of this paper, it is still a crucial part of the collision detection pipeline.
Elseberg et al.~\cite{6102102} proposed data structures and algorithms to efficiently construct octrees from point cloud data, such as from laser scanners on robots.
They also proposed an efficient octree to store and compress 3D data without precision loss, especially on large point cloud datasets~\cite{ELSEBERG201376}.
Su et al.~\cite{SU201659} explored an algorithm to perform segmentation on LiDAR point cloud data using octree decomposition.
Hao et al.~\cite{rs70809682} proposed a method to construct octrees from LiDAR data that specializes in detecting changes to buildings and trees in an urban environment.
More recently, OctoMap-RT~\cite{min2023octomap} and OctoCache~\cite{chen2025octocache} accelerate the OctoMap~\cite{hornung2013octomap} octree representation build and update operations, while others like i-Octree~\cite{zhu2024octree} target SLAM.
These works ensure that the octree representation matches the current state of the environment and can easily be used in collision detection.
BVH trees are also commonly used to represent the environment, with various construction algorithms such as LBVH~\cite{lauterbach2009fast}, PLOC~\cite{meister2017parallel}, H-PLOC~\cite{hploc}, PLOC++~\cite{benthin2022ploc++}, and CWBVH~\cite{ylitie2017efficient} that may offer better performance for some robotics tasks.

\section{Conclusion}
\label{sec:conclusion}

In this paper, we present \roboRT, a robotics-focused hardware unit integrated into a RoboGPU architecture to accelerate collision detection and other robotics tasks.
Collision detection is crucial in neural planners but requires performance that can be difficult to achieve on GPUs alone.
We propose simple yet effective architectural modifications that convert RTAs into \roboRT to be useful in robotics.
\roboRT enhances the GPU and achieves \speedupCollision speedup over the state-of-the-art Mochi~\cite{mandarapu2024mochi} implementation for collision detection, while also accelerating M$\pi$Net~\cite{mpinet} by \speedupMpiNet and Monte Carlo Localization from the RoWild~\cite{rowild} by 1.1$\times$.
\roboRT explores an architecture that enables the first step towards safe deployment of neural planners, but use on real robotic systems will likely require further improvements before any safety guarantees can be made.
Our results show that GPUs with specialized cores offers a flexible and efficient platform that can adapt to the rapidly evolving landscape of robotics while meeting performance demands.


\section*{Acknowledgments}
The authors would like to thank the anonymous reviewers for their helpful comments.
This work was supported by the Natural Sciences and Engineering Research Council of Canada (NSERC). 
Tor Aamodt has recently served as a consultant for NVIDIA Corp. 
Special mention to Alexander Zheng.


\bibliographystyle{IEEEtran}
\bibliography{paper}

\end{document}